\documentclass[3p, number,sort&compress,times,fleqn]{elsarticle}      

\usepackage{amsmath}
\usepackage{amssymb}
\usepackage{mathtools}
\usepackage[colorlinks = true,
linkcolor = blue,
urlcolor  = blue,
citecolor = blue,
anchorcolor = blue]{hyperref}

\usepackage{subfig}
\usepackage{graphicx}
\usepackage[labelfont=bf, justification=justified]{caption}

\usepackage{float}

\usepackage[ruled]{algorithm2e}
\usepackage[noend]{algpseudocode}

\usepackage{csml}

\graphicspath{{./figs/}}
\usepackage{url}

\begin{document}
	
\begin{frontmatter}
	
\title{Topologically robust CAD model generation for structural optimisation}

\author{Ge Yin}
\author{Xiao Xiao}
\author{Fehmi Cirak\corref{cor1}}
\ead{f.cirak@eng.cam.ac.uk}

\cortext[cor1]{Corresponding author}

\address{Department of Engineering, University of Cambridge, Trumpington Street, Cambridge, CB2 1PZ, UK}

\begin{abstract}
Computer-aided design (CAD) models play a crucial role in the design, manufacturing and maintenance of products. Therefore, the mesh-based finite element descriptions common in structural optimisation must be first translated into CAD models. Currently, this can at best be performed semi-manually. We propose a fully automated and topologically accurate approach to synthesise a structurally-sound parametric CAD model from topology optimised finite element models. Our solution is to first convert the topology optimised structure into a spatial frame structure and then to regenerate it in a CAD system using standard constructive solid geometry (CSG) operations. The obtained parametric CAD models are compact, that is, have as few as possible geometric parameters, which makes them ideal for editing and further processing within a CAD system. The critical task of converting the topology optimised structure into an optimal spatial frame structure is accomplished in several steps. We first generate from the topology optimised voxel model a one-voxel-wide voxel chain model using a topology-preserving skeletonisation algorithm from digital topology. The weighted undirected graph defined by the voxel chain model yields a spatial frame structure after processing it with standard graph algorithms. Subsequently, we optimise the cross-sections and layout of the frame members to recover its optimality, which may have been compromised during the conversion process. At last, we generate the obtained frame structure in a CAD system by repeatedly combining primitive solids, like cylinders and spheres, using boolean operations. The resulting solid model is a boundary representation (B-Rep) consisting of trimmed non-uniform rational B-spline (NURBS) curves and surfaces. 
\end{abstract}
	
\begin{keyword}
topology optimisation, computer-aided design, digital topology, homotopic skeletonisation, CSG tree
\end{keyword}

\end{frontmatter}

%
\section{Introduction \label{sec:intro}}
%
The application of structural optimisation in industrial design requires the optimised geometries to be converted into CAD models. This is necessary because CAD systems are today an integral part of  most product development processes. The prevalent parametric CAD systems are based on boundary representation (B-rep) techniques and trimmed NURBS (non-uniform rational B-splines). Geometries are created using constructive solid geometry (CSG) by recursively combining primitive shapes, such as  cylinders, cubes etc. using boolean operations. This  recursive process is stored in the form of a CSG tree.  Although the input to most structural optimisation is a CAD geometry, their output is usually a finite element mesh of the optimised geometry. The finite element meshes consist of too many elements to be processed and edited in a CAD system. For a geometry to be editable in a CAD system, a compact representation in the form of a CSG tree is needed. The need to edit the optimised part geometry arises when additional geometric features have to be added or when the part is to be combined with other components into a functional product. Moreover, in industrial practice there are usually, in addition to structural efficiency and robustness, many equally important, often not explicitly quantified, design requirements, which require the geometry to be edited after optimisation.

\begin{figure}[tb]
\centering
	\subfloat[][Design space and loading \label{fig:rockerArmA}] {
		\includegraphics[width=0.27\textwidth]{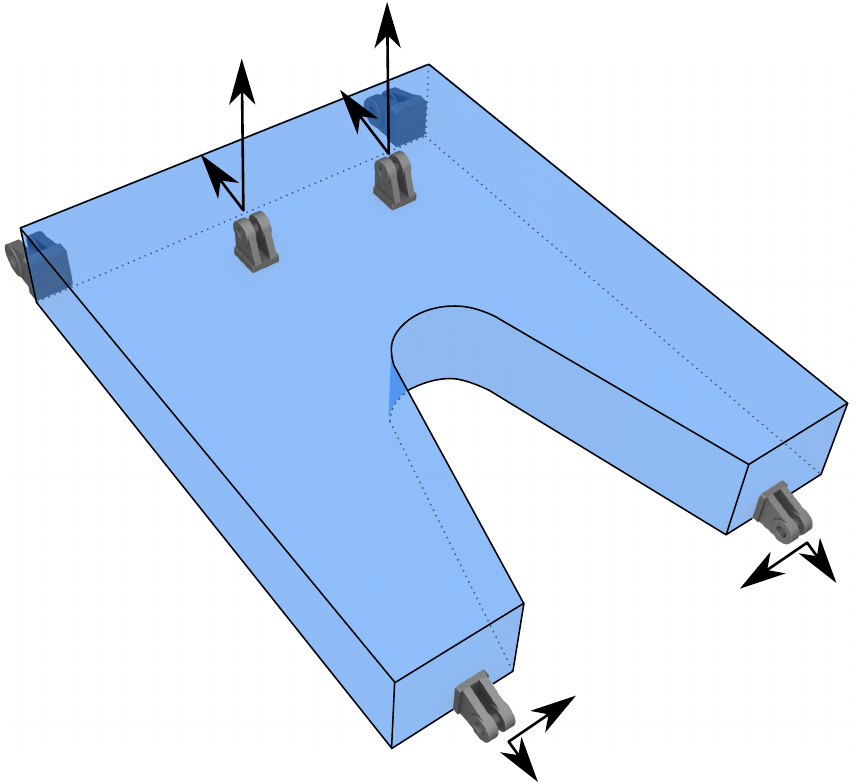} }
\hspace{0.035\textwidth}
	\subfloat[][Topology optimised geometry \label{fig:rockerArmB}] {
		\includegraphics[width=0.27\textwidth]{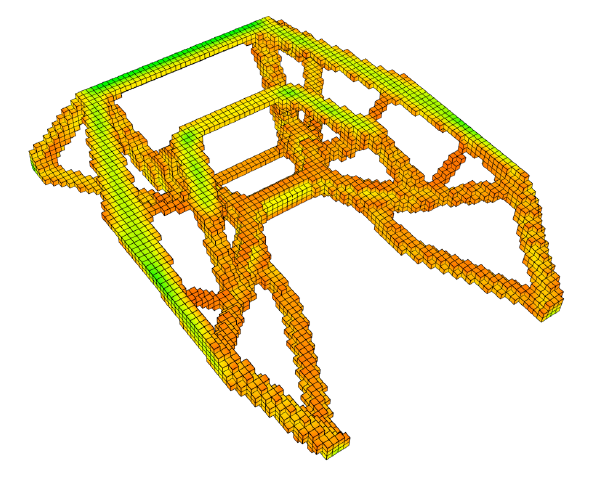}  }
\hspace{0.035\textwidth} 
	\subfloat[][Homotopically thinned voxel chain model \label{fig:rockerArmC}] { 
		\includegraphics[width=0.27\textwidth]{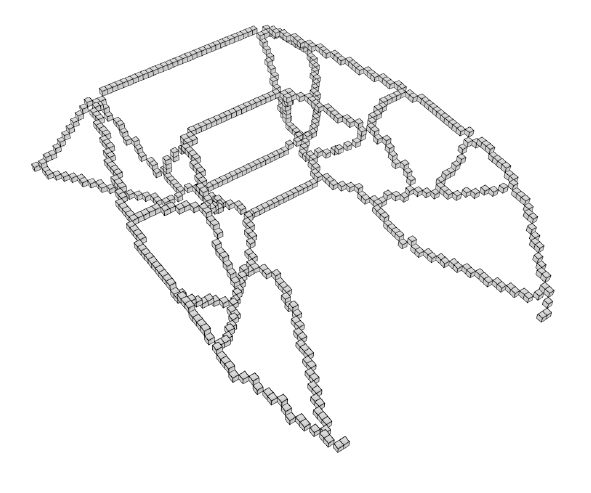} }
\\[-0.5em]
	\subfloat[][Spatial frame model \label{fig:rockerArmD}] { 
		\includegraphics[width=0.27\textwidth]{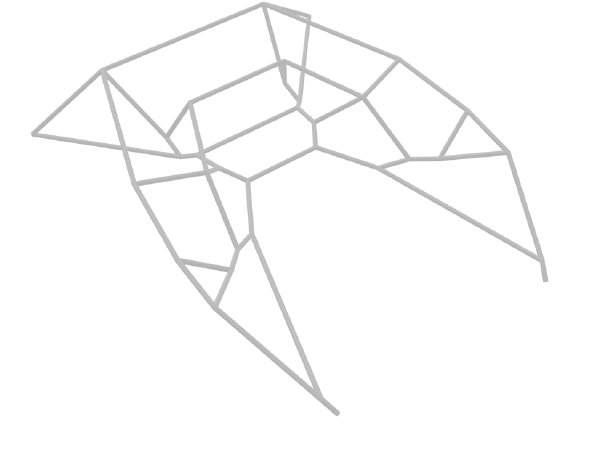} }
\hspace{0.035\textwidth} 
	\subfloat[][CAD model generated by recursively combining primitive solids  \label{fig:rockerArmE}] {
		\includegraphics[width=0.27\textwidth]{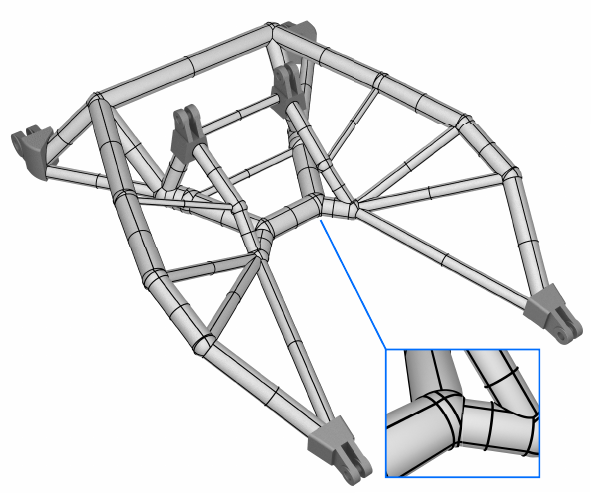}  }
\hspace{0.035\textwidth} 
	\subfloat[][CAD model generated by combining primitive solids and blended surfaces \label{fig:rockerArmF}] {
		\includegraphics[width=0.27\textwidth]{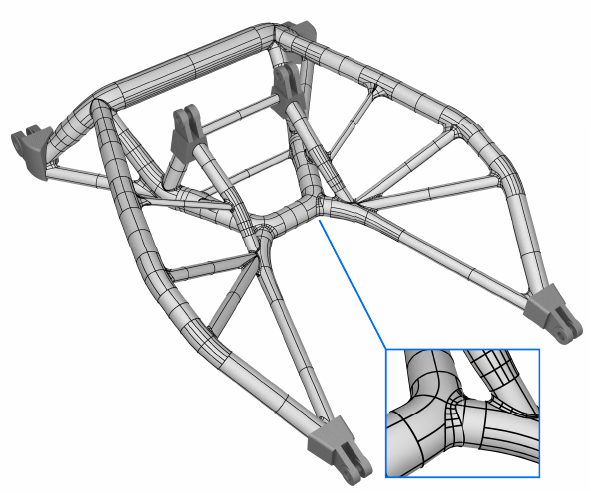}  }
	\caption{CAD model generation workflow for a topology optimised rocker arm. The rocker arm is a lever mechanism which pivots around the two of the joints shown in (a) when a force is applied at the other four joints. In (e) and (f) the six detailed joint designs, with each containing a roller bearing, have been added in a CAD system after model generation. 
	\label{fig:rockerArm}}
\end{figure}

The fully automated process we propose  can  robustly generate a compact CSG tree representation of topology optimised geometries and convert them to CAD models. The entire workflow is illustrated with the help of the topology optimisation of the rocker arm shown in Figure~\ref{fig:rockerArm}.  For topology optimisation we use a standard density based SIMP (solid isotropic material with penalisation) approach on a structured hexahedral finite element mesh referred to as a voxel mesh, see e.g.~\cite{bendsoe1989optimal, sigmund200199, bendsoe2003topology}. The optimised density field splits the voxels after thresholding into the subsets solid and void,  which gives a three-dimensional binary image with two grey levels. Subsequently, we extract from the binary image a voxel chain skeleton  using a homotopic, i.e. topology-preserving, skeletonisation algorithm from digital topology~\cite{lee1994building}.  The skeleton has  the same number of connected components, handles and cavities as the three-dimensional voxel model. Evidently, topology preservation is critical in structural applications because a change of topology may disrupt the load paths discovered during optimisation.  The voxel chain model defines a weighted undirected graph which is processed with graph algorithms to obtain a spatial frame structure. Each of the members of the frame structure is a beam and the beams are rigidly connected at the joints. During the conversion of the voxel model  to a spatial frame structure the optimality of the structure is usually compromised. This can, however, be recovered after applying a few steps of size and layout optimisation to the spatial frame structure. Size optimisation adjusts the cross-sections of the members and layout optimisation adjusts the coordinates of the joints. The very compact representation which we generate from the original voxel model consists of the connectivity of the frame structure, its joint coordinates and the member cross-sections. These are used to create a binary CSG tree and a solid model of the spatial frame structure in a CAD system. Figures~\ref{fig:rockerArmE} and~\ref{fig:rockerArmF} show two different solid models generated in a CAD system. In the first model the cylindrical members are connected to spherical joints. In the second model the members are still cylinders, which are, however, smoothly blended at the joints. With the process presented in this paper, the first model is generated in a fully automated fashion while the second requires some user intervention. Note although  we used a density based approach for topology optimisation, it is straightforward to apply the proposed approach to the more recent level-set based methods~\cite{wang2003level,Allaire:2004aa} or the historical homogenisation method~\cite{bendsoe1988generating}, see also~\cite{sigmund2013topology} for an overview on optimisation approaches. 

There are only a very limited number of  structural optimisation approaches which arrive at a geometry in the form of a compact parametric CAD model. 
In more recent topology optimisation formulations the placement of a finite number of geometric primitives, such as rectangles or cuboids, embedded in a larger design domain is considered~\cite{guo2014doing,norato2015geometry,zhou2016feature,xie2018new}. The primitives are allowed to freely move within the design domain usually discretised with a fixed-grid finite element approach. The ensuing optimisation problem is inherently discrete and is converted into a continuous problem by careful regularisation. Although these techniques can provide a compact CAD model they are usually incompatible with the commercially prevalent density based topology optimisation implementations and have so far been applied mostly to 2D problems. There are also  other approaches primarily intended for shape optimisation. 
For instance, it is  quite common to use the parameters of a CAD or a reconstructed CAD-like spline model as geometric design variables in shape optimisation, see e.g.~\cite{Braibant:1984aa, Bletzinger:1991aa, robinson2012optimizing, han2014adaptive}. Such techniques are especially appealing in an isogeometric analysis context when the same basis functions are used for geometry description and finite element analysis~\cite{Cirak:2002aa, Wall:2008aa, Kiendl:2009aa, bandara2016shape, herrema2017framework, bandara:2018}. The compact spline representation of the optimised geometry can in principle be imported into a CAD system. In the past, there has been some work in topology optimisation on generating a spline representation based on a given optimised finite element geometry. A common approach is to fit the density isocontours obtained with topology optimisation with a spline curve in 2D or a spline surface in 3D~\cite{maute1995adaptive, lin2000automated, hsu2001interpreting}. Especially in 3D, these techniques are not robust as they may require the solution of a nonlinear least squares problem and are hard to automate due to the need to manually position the  control vertices of the splines. Possibly, therefore,  they had no noteworthy impact on commercial software despite being developed around two decades ago. The use of  skeletonisation, or thinning, for compact geometric representations has  earlier been pioneered in Bremicker et al.~\cite{bremicker1991integrated} for 2D and very recently in Cuilli\`ere~\cite{cuilliere2018automatic} for 3D. While in~\cite{bremicker1991integrated} the aim was not to arrive at a parametric CAD model, skeletonisation algorithms from digital topology were used to extract a truss structure, which was subsequently size and layout optimised. Homotopic skeletonisation in 3D is  substantially more challenging than in 2D, and elimination of possible mechanisms in a spatial truss structure is far from trivial~\cite{pellegrino1993structural}. And so, in our approach, we convert the voxel chain model into a spatial frame structure (with rigidly connected joints) which intrinsically does not exhibit mechanisms. In~\cite{cuilliere2018automatic} a curve skeleton is extracted  from the surface mesh representing the density isocontour of the optimised geometry using the skeleton extraction method proposed by Au et al. ~\cite{au2008skeleton}.  This specific extraction technique is rather elaborate as it builds on the successive contraction of a given surface mesh using Laplacian smoothing. The obtained skeleton and estimated cross-sections are used to generate a CAD model with no further postprocessing. It is not taken into account that skeletonisation may have impaired the optimality of the structure.  A more manufacturing oriented overview of compact parametric representations for topology optimised geometries can be found in the recent review~\cite{liu2016survey}.

An essential component of the proposed CAD model generation process is skeletonisation, which is an active research area with applications in computer graphics, animation and volumetric image processing, see the comprehensive reviews~\cite{cornea2007curve, siddiqi2008medial, saha2016survey, tagliasacchi20163d, saha2017skeletonization}. The curve-skeleton, briefly the skeleton, of a 3D object is closely related to its medial surface, or surface-skeleton. In finite elements medial surfaces are known from mesh generation applications, see e.g.~\cite{sheehy1996shape}.   Informally, the medial surface is the set of all points that have two or more closest points  on a 3D object's surface. That is, a sphere centred on the medial surface touches the object's surface at two or more points.  In contrast to the medial surface, the skeleton of a 3D object has no rigorous definition. It is expected that the skeleton captures the essential topology of the object and is centred, i.e. lies on or close to the medial surface. There are  many algorithms for determining the skeleton of an object starting from different types of geometry representations, such as implicit signed distances, polygonal surfaces etc. In topology optimisation, it is expedient to assume that the  object is given as a voxelised binary image consisting of solid and void voxels. Digital topology provides a rigorous basis to study the topological properties of binary images consisting of pixels or voxels. An accessible introduction to digital topology can be found in the review~\cite{kong1989digital} and the textbook~\cite{klette2004digital}.  Homotopic skeletonisation algorithms based on digital topology can extract a one-voxel-wide skeleton with the same topology as the binary image, i.e. the same number of connected components, handles and cavities. See the excellent reviews~\cite{kong1989digital, lam1992thinning} for early papers and~\cite{saha2016survey, saha2017skeletonization} for more recent papers on skeletonisation.  Most of these methods start from the domain boundaries and proceed by iteratively removing voxels one by one that do not alter the topology of the object. When the algorithms terminate, the entire structure is represented by a network of one-voxel-wide chains. It is sufficient to inspect only a voxel's immediate neighbourhood consisting of 3$\times$3$\times$3 voxels to decide whether that voxel can be removed. The possible states of such a small cluster can effectively be encoded in a look-up table~\cite{lobregt1980three,lee1994building}. Crucially, the skeletonisation process does not rely on any floating-point operations, which makes it exceedingly robust. Recent research on skeletonisation focuses on parallelisation~\cite{ma2002three,lohou20053d} and improving the memory usage~\cite{yan2018voxel} of the algorithms. Skeletonisation is however far less computing intensive than finite element analysis so that advanced skeletonisation algorithms are of limited relevance for the present work. Our approach is based on Lee et al.~\cite{lee1994building} which is one of the first provably homotopy preserving skeletonisation algorithms. Its implementation is also discussed in the more recent papers~\cite{homann2007,post2016fast}.

The outline of this paper is as follows. In Section~\ref{sec:structOpt}, the standard density based topology optimisation and the size and layout optimisation of spatial frames are briefly reviewed. Subsequently,  in Section~\ref{sec:digTopSkeleton}  relevant aspects of digital topology are introduced and the implemented specific skeletonisation algorithm is described. The robustness and runtime of our implementation are studied with a relatively complex quadcopter frame geometry. The postprocessing of the obtained voxel chain model with graph algorithms and the generation of first  a frame structure and then a CAD model are discussed in Section~\ref{sec:extractionAndCovnversion}. The entire optimisation and geometry conversion process is summarised in Section~\ref{sec:overall}. The application of the proposed approach to  a standard 2D benchmark example from topology optimisation and more complex 3D examples is illustrated in Section~\ref{sec:examples}. We study in particular the change of the compliance cost function during the conversion of the optimised voxel model to a CAD model.

%
\section{Review of structural optimisation \label{sec:structOpt}}
%
In this Section, we briefly review the standard density based topology optimisation using the SIMP method and the size and layout optimisation of spatial frame structures. In all cases, the cost function is the compliance, and the discussion is restricted to aspects relevant to this paper. For further details see, e.g., the monographs~\cite{bendsoe2003topology, choi2006structural}. 
%
\subsection{Topology optimisation of solids \label{sec:topOpt}}
%

The topology optimisation problem for finite element discretised solids reads 
\begin{subequations} \label{eq:topOpt}
\begin{align}
	\text{minimise } \quad  & J( \vec \rho) = \vec f \cdot \vec u (\vec \rho)  \label{eq:topOptA}\\
	\text{subject to} \quad  & \vec K (\vec \rho) \vec u  = \vec f  \label{eq:topOptB}\\
	& \frac{V(\vec \rho)}{\overline V} \le V_f \label{eq:topOptC} \\
	& \vec 0 \le \vec \rho \le  \vec 1  \, ,
\end{align}
\end{subequations}
where~$J(\vec \rho)$ is the compliance cost function, \vec{\rho} is the vector of relative element densities, $\vec{u}$ is the displacement vector, $\vec K$ is the global stiffness matrix,  \vec{f} is the global external force vector, $V(\vec \rho)$ is the material volume,  $\overline V$ is the design domain volume and $V_f$ is the scalar prescribed volume fraction. The relative density of each element is constrained to be~\mbox{$0 \le \rho_i \le 1$}  with $i \in \{1, \dotsc, n_e \}$ and the number of elements is $n_e$. 

As usual, the global stiffness matrix~$\vec K$ and vector~\vec{f} are assembled from the~$n_e$ element contributions~$\vec K_i$ and~$\vec f_i$ in the mesh.  In this paper, we discretise the design domain always with a structured grid and hexahedral linear elements.  In each element~$i$ the material is isotropic and homogeneous, and the Young's modulus $E$ is penalised depending on the relative density~$\rho_i$ with 
\begin{equation} \label{eq:younPenalised}
	E  (\rho_i) = E_{\text{min}} + \rho_i^p (\overline{E}- E_{\text{min}}) \, ,
\end{equation}
where~$\overline{E}$ is the prescribed Young's modulus of the solid material, $p$ and $E_{\text{min}} $ are two algorithmic parameters. The penalisation parameter~$p \ge 3$ ensures that elements with densities close to~$\rho_i = 0$ (void) and~$\rho_i = 1$ (solid) are preferred. The small Young's modulus  $E_{\text{min}} \approx  10^{-9}$ of the void material prevents ill-conditioning of the global stiffness matrix when~$\rho_i = 0$. Each element stiffness matrix~$\vec K_i$ is computed using the corresponding Young's modulus~$E(\rho_i)$  with the relative density~$\rho_i$, which is constant within an  element. 

Furthermore, the topology optimisation problem~\eqref{eq:topOpt} is regularised by filtering the element densities~$\rho_i$ to prevent checker-board instabilities and mesh dependency of the solution. This is accomplished by convolving the element densities with the kernel function
\begin{equation}
	H(\vec x_i, \, \vec x_j) = 
	\begin{cases} 
		R - \text{dist}(\vec x_i, \, \vec x_j)  &  \text{if} \, \,   \text{dist } (\vec x_i, \,  \vec x_j) \le R  \\
		0 &  \quad \text{else}  
		\end{cases} \, ,
\end{equation} 
which depends on the coordinates of the centroids~$\vec x_i$ and $\vec x_j$ of the elements~$i$ and~$j$, and the prescribed filter length~$R$.  With~$H(\vec x_i,\,  \vec x_j)$ at hand the filtered densities are given by
\begin{equation} \label{eq:filter}
	\hat \rho_i  = \frac{\sum_j H (\vec x_i, \, \vec x_j)   \rho_j} { \sum_j H (\vec x_i, \, \vec x_j)  } 
\end{equation}
assuming that all elements have the same volume.

For gradient-based optimisation the derivatives of the cost and constraint functions in~\eqref{eq:topOpt} with respect to the relative densities~$ \rho_i$ are needed. In the following  the  relative densities~$\rho_i$ in the topology optimisation problem~\eqref{eq:topOpt}  are replaced with the filtered relative densities~$\hat {\rho}_i$. The derivative, or sensitivity, of the compliance cost function~\eqref{eq:topOptA} reads
\begin{equation} \label{eq:sensitivity}
	\frac{\partial J (\hat {\vec \rho})}{ \partial  \rho_i} =  \sum_j  \vec f \cdot \frac{\partial \vec u  (\hat{\vec \rho}) }{\partial \hat \rho_j} \frac{\partial \hat {\rho}_j}{\partial \rho_i} \, .
\end{equation}
Differentiating and rearranging the equilibrium constraint~\eqref{eq:topOptB} gives 
\begin{equation}
	\frac{\partial \vec u (\hat{\vec \rho})}{\partial \hat \rho_j} = -  \vec K^{-1}(\hat {\vec \rho}) \frac{\partial \vec K (\hat {\vec \rho})}{\partial \hat \rho_j} \vec u (\hat{\vec \rho})
\end{equation}
and the derivative of the filtered densities~\eqref{eq:filter} is
\begin{equation}
	\frac{\partial \hat \rho_j}{\partial \rho_i} =  \frac{H (\vec x_j , \vec x_i)}{\sum_k H (\vec x_j, \, \vec x_k)}  \, .
\end{equation}  
Introducing both derivatives into~\eqref{eq:sensitivity} yields
\begin{equation} \label{eq:sensitivity2}
	\frac{\partial J (\hat {\vec \rho})}{ \partial  \rho_i} = - \sum_j  \vec u (\hat{\vec \rho})  \cdot \frac{\partial \vec K (\hat {\vec \rho})}{\partial \hat \rho_j} \vec u (\hat{\vec \rho}) \frac{H (\vec x_j , \vec x_i)}{\sum_k H (\vec x_j, \, \vec x_k)}     \, ,
\end{equation}
where the stiffness matrix derivative is to be assembled from element contributions considering the penalised Young's modulus~\eqref{eq:younPenalised}. 

The derivative of the volume constraint~\eqref{eq:topOptC}  is  obtained analogously with
\begin{equation} \label{eq:volSensitivity}
	\frac{\partial V (\hat {\vec \rho})}{ \partial  \rho_i} = \sum_j \hat \rho_j \frac{V}{n_e} \frac{H (\vec x_j , \vec x_i)}{\sum_k H (\vec x_j, \, \vec x_k)}     \, . 
\end{equation}
It is assumed here that all the elements are of the same size, namely the design volume~$\overline V$ divided by the total number of elements~$n_e$. 

Finally, with the obtained derivative of the compliance cost function~\eqref{eq:sensitivity2} and the volume constraint~\eqref{eq:volSensitivity} the optimised density distribution is determined iteratively with the optimality criteria method, see e.g.~\cite{bendsoe2003topology}.

%
\subsection{Size and layout optimisation of frames \label{sec:frameOpt}}
%
We consider spatial frame structures consisting of straight beam members connected by joints that can transfer forces and moments. The members can deform by stretching, bending and torsion and are modelled as classical Timoshenko beams, see~\ref{sec:appendix}. Without loss of generality, size and layout optimisation can be posed as an iterative sequential optimisation problem.  In each optimisation step, either the size, i.e. cross-section area, or layout, i.e. joint coordinates, of the members is optimised. 

Both the size and layout optimisation problems have the same structure as topology optimisation~\eqref{eq:topOpt}, namely, 
\begin{subequations} \label{eq:frameOpt}
\begin{align}
	\text{minimise } \quad  & J(\vec s) = \vec u(\vec s) \cdot \vec K(\vec s) \vec u (\vec s) = \vec f \cdot \vec u (\vec s) \label{eq:frameOptA}\\
	\text{subject to} \quad  & \vec K (\vec s) \vec u (\vec s) = \vec f  \label{eq:frameOptB}\\
	& \frac{V(\vec s)}{\overline V} \le V_f \label{eq:frameOptC} \\
	& \vec s_l \le \vec s \le  \vec s_u  \, .
\end{align}
\end{subequations}
The vector of design variables $\vec{s}$ refers either to cross-section areas in case of size optimisation or joint coordinates in case of layout optimisation. The two bounds~$\vec s_l$ and~$\vec s_u$ have different interpretations in the two cases.

The frame consists of~$n_e$ beam elements with the element stiffness matrices~$\vec K_i$ and vectors~$\vec f_i$.  Each stiffness matrix~$\vec K_i$  is obtained from a local stiffness matrix~$\vec K_i^l$ formulated in a coordinate system attached to the element with the index~$i$. The local matrix~$\vec K_i^l \in \mathbb R^{12 \times 12}$ with the displacements and rotations of the two end nodes of the beam element as degrees of freedom is given in \ref{sec:appendix}. In the local $x^l_i$, $y^l_i$ and $z^l_i$ coordinate system the beam axis is assumed to be aligned with the $ x^l_i$ axis. The matrix~$K_i^l$ is transformed to the global $x$, $y$ and $z$ coordinate system according to
\begin{equation} \label{eq:elmStiffness}
	\vec{K}_i = \vec{\Lambda}_i  \vec{K}_i^l \vec{\Lambda}_i^\trans 
\end{equation}
with the block-diagonal transformation matrix 
\begin{equation}
	\vec{\Lambda}_i = \diag ( \vec{\lambda}_i, \, \vec{\lambda}_i,  \, \vec{\lambda}_i,  \, \vec{\lambda}_i )\, .
\end{equation}
The rotation matrix~$\vec \lambda_i \in \mathbb R^{3\times3}$ can be chosen with 
\begin{equation}
	\vec{\lambda}_i = 
	\begin{pmatrix}
		\cos\alpha_i \cos \beta_i & -\sin \alpha_i  & \cos \alpha_i \sin \beta_i   \\
		\sin\alpha_i \cos \beta_i  & \phantom{+} \cos \alpha_i & \sin \alpha_i \sin \beta_i  \\
		- \sin \beta_i  &
		0 & \cos \beta_i
	\end{pmatrix} \, ,
\end{equation}
which is composed of the two elemental rotations~$\alpha_i$ around the $z_i^l$  axis and $\beta_i$ around  $y_i^l$ axis.

Similar as in topology optimisation, the derivative, or  sensitivity, of the compliance cost function~\eqref{eq:frameOptA} reads
\begin{equation} \label{eq:complianceDerivatives}
\frac{\partial J(\vec{s})}{\partial s_i} =
-\vec{u} (\vec{s}) \cdot \frac{\partial\vec{K}(\vec{s})}{\partial s_i}\vec{u}(\vec{s}) =  -\sum_{j = 1}^{n_e}\vec{u}_j (\vec{s}) \cdot \frac{\partial\vec{K}_j (\vec{s})}{\partial s_i}\vec{u}_j (\vec{s}) \, ,
\end{equation}
where $\vec{u}_j$ is the vector of nodal displacements and rotations of the element~$j$. In size optimisation the derivative of the element stiffness matrix~\eqref{eq:elmStiffness} is given by 
\begin{equation}
\frac{\partial \vec{K}_i(\vec{s})}{\partial s_j} = \vec{\Lambda}_i \frac{\partial{\vec{K}}_i^l}{\partial s_j}\vec{\Lambda}_i^\trans \, ,
\end{equation}
whereas in layout optimisation it is
\begin{equation}
\frac{\partial \vec{K}_i(\vec{s})}{\partial s_j} = 
\frac{\partial\vec{\Lambda}_i}{\partial s_j}  \vec K_i^l \vec{\Lambda}_i^\trans +
\vec{\Lambda}_i \frac{\partial \vec K_i^l}{\partial s_j}\vec{\Lambda}_i^\trans +
\vec{\Lambda}_i  \vec K_i^l \frac{\partial\vec{\Lambda}_i^\trans}{\partial s_j} \, .
\end{equation}
%

%
\section{Digital topology and skeletonisation \label{sec:digTopSkeleton}}
%
The input to skeletonisation is a structured hexahedral finite element grid with all the vertices within the domain having eight incident hexahedra. For digital topology only the connectivity of the mesh is relevant whereas the coordinates of the grid nodes  can be arbitrary so that in the following hexahedron and voxel are used interchangeably.  The hexahedral grid encodes the binary voxel model of the topology optimised structure.  The binary voxel model is obtained by denoting voxels with a density~$\rho_i$ above the threshold~$\eta$ as {\em solid} and the remaining ones as {\em empty (void)}. 
Hence, the set of  all voxels~\mbox{$ v \in \set V =  \set V_s \cup \set V_e$} is composed of the solid voxels~$ \set V_s$ and empty voxels~$ \set V_e$. Skeletonisation aims to reassign voxels one by one from~$ \set V_s$ to~$ \set V_e$ until the solid is represented by a network of one-voxel-wide  chains. The decision about which voxels to reassign is guided by digital topology and proceeds starting from the voxels at the border between~$\set V_s$ and~$\set V_e$.

%
\subsection{Review of digital topology \label{sec:digTop}}
%

\begin{figure}
\centering
\subfloat[][
$  \chi( \set M_s) = 1 ;  \, \,   \chi(\partial \set M_s)  = 2 $
] {
\includegraphics[scale=0.5]{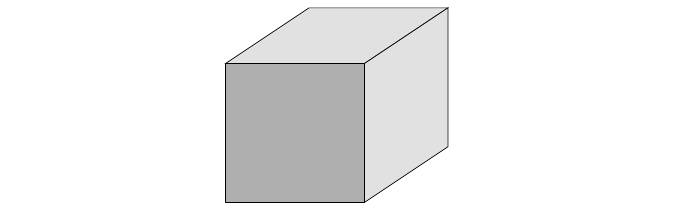} }
\hspace{0.1\textwidth}
\subfloat[][
$ \chi( \set M_s) = 1  ;  \, \, \chi(\partial \set M_s)  =2 $] {
\includegraphics[scale=0.5]{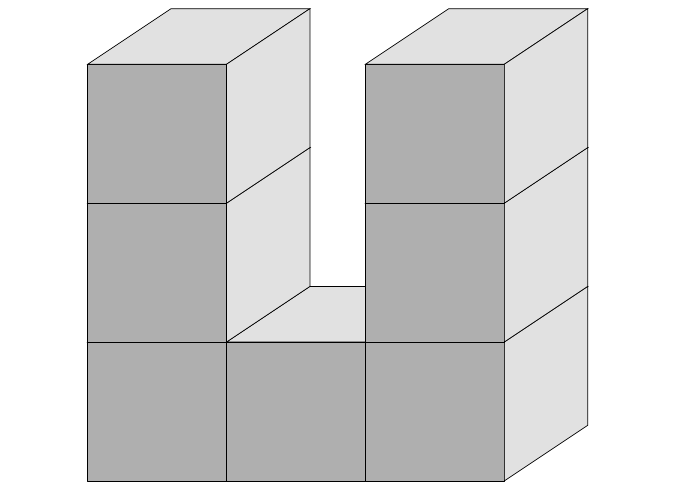}  }
\hspace{0.1\textwidth} 
\subfloat[][
$\chi( \set M_s) = 0  ;  \, \, \chi(\partial \set M_s) = 0$] { 
\includegraphics[scale=0.5]{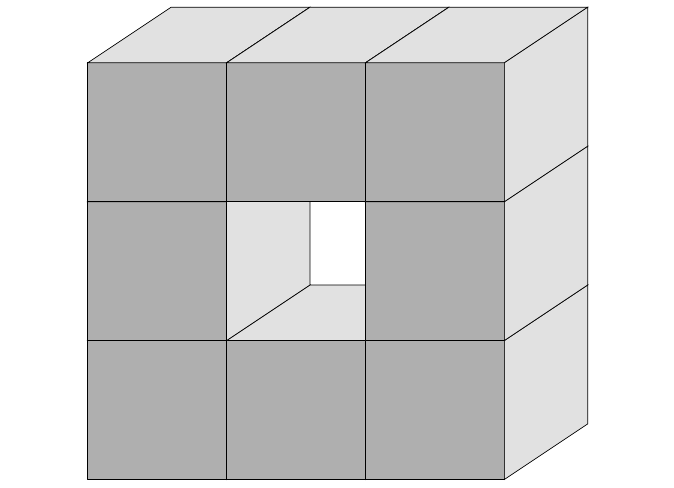} }
\caption{Euler characteristics $\chi( \set M_s)$ and $ \chi(\partial \set M_s)$ for three different meshes. The number of different entities for the volume mesh in (a) are $n_0=8$, $n_1=12$, $n_2=6$ and $n_3=1$; for the mesh in (b) they are $n_0=32$, $n_1=60$, $n_2=36$ and $n_3=7$;  and, for the mesh in (c) they are $n_0=32$, $n_1=64$, $n_2=40$ and $n_3=8$. } \label{fig:voxelsEuler}
\end{figure}

To retain the topology of the voxel model, its Euler characteristic is considered. The solid voxels in~$ \set V_s$ and empty voxels in~$ \set V_e$ yield two hexahedral volume meshes, or more abstractly cell complexes, which will be in the following denoted with~$\set M_s  = \set M (\set V_s)$ and~$\set M_e = \set M (\set V_e)$. The volume mesh of a voxel set consists of the corresponding set of the vertices, edges, faces and voxels. It is also useful to define quadrilateral surface meshes~$ \partial \set M_s$ and~$ \partial \set M_e$ formed by the boundaries of~$\set M_s$ and~$\set M_e$, respectively.  The Euler characteristic of the voxel model can be determined either from the volume mesh~$ \set M_s$ with   
\begin{equation} \label{eq:eulerVol}
	\chi  (\set M_s) = n_0 (\set M_s)  - n_1 (\set M_s) + n_2 (\set M_s) - n_3 (\set M_s) \, ,
\end{equation}
or from the surface mesh~$\partial \set M_s$ with 
\begin{equation} \label{eq:eulerSurf}
	\chi  (\partial \set M_s) = n_0 (\partial \set M_s) - n_1 (\partial \set M_s)+ n_2 (\partial \set M_s) \, ,
\end{equation}
where~$n_0$~denotes the number of vertices, $n_1$~the number of edges, $n_2$~the number of the faces and $n_3$~the number of hexahedrons in~$\set M_s$ or~$ \partial \set M_s$.  In Figure~\ref{fig:voxelsEuler} the Euler characteristics of three voxel meshes are given.  Notice that there is also the relation
\begin{equation}
	\chi(\partial \set M_s) = 2 \chi (\set M_s)
\end{equation}
 between the Euler characteristics of the surface and volume meshes. The Euler characteristic of a  voxel model is related to the total number of separate objects (connected components)~$o(\set M_s)$, handles (tunnels)~$h(\set M_s)$ and cavities~$c(\set M_s)$  in the entire model according to the formula
\begin{equation} \label{eq:eulerGlobal}
		\chi (\set M_s) = o(\set M_s) - h(\set M_s) + c(\set M_s) \, , 
\end{equation}
see, e.g., textbooks~\cite{hatcher2005algebraic, crossley2006essential} on topology. For instance, the Euler characteristic $\chi(\set M_s)$ of a sphere is~$1$, of a torus is~$0$, of a double torus is~$-1$ and so on, compare also Figure~\ref{fig:voxelsEuler}. Hence, the total number of different entities in the mesh and the described shape are intrinsically connected. 

\begin{figure}
	\centering
	\subfloat[][6-neighbours ($\set N_{6}(v)$)] {						
		\includegraphics[scale=0.5]{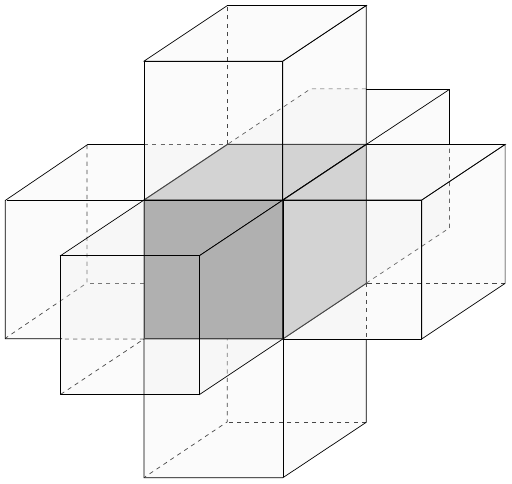}
		\label{fig:6neighbourship}}
\hspace{0.05\textwidth}
	\subfloat[][26-neighbours ($\set N_{26}(v)$)]{						
		\includegraphics[scale=0.5]{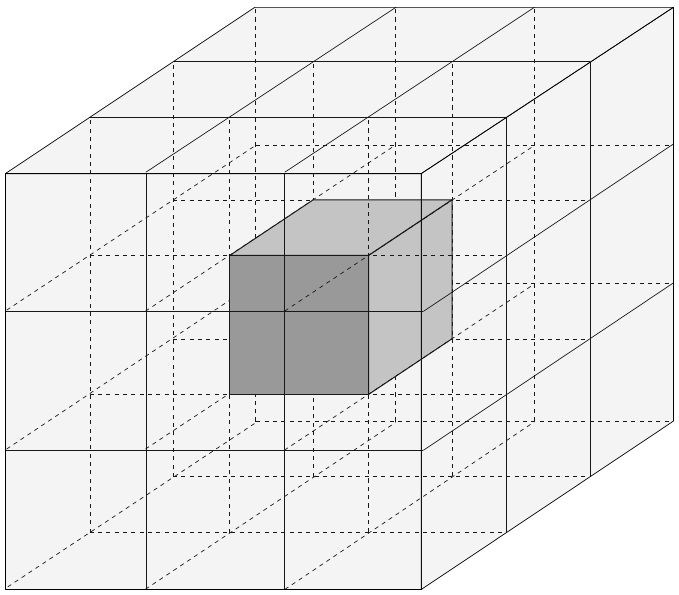}
		\label{fig:26neighbourship}}
\hspace{0.05\textwidth}
	\subfloat[][Not 6-neighbours, but 26-neighbours]{	
		\includegraphics[scale=0.5]{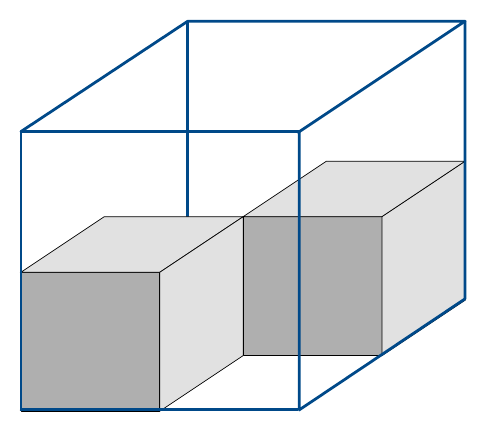}
		\label{fig:touchingVoxels_1}}
\hspace{0.05\textwidth}
	\subfloat[][Not 6-neighbours, but 26-neighbours]{						
		\includegraphics[scale=0.5]{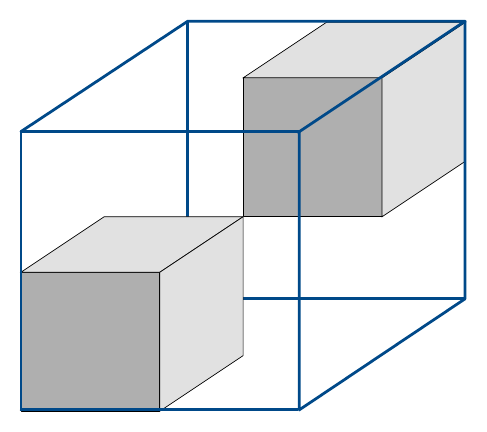}
		\label{fig:touchingVoxels_2}}
\caption{Voxel neighbourship definitions. In (a) and (b) the light shaded voxels are the~$\set N_{6}(v)$ and~$\set N_{26}(v)$ neighbours of the dark shaded voxel~$v$  at the centre. The two voxels in (c) and  (d) are each others 26-neighbour but not 6-neighbour. } \label{fig:voxelNeighborship}
\end{figure}

In determining the Euler characteristic of a voxel model, two different types of neighbourship definitions are possible, see Figure~\ref{fig:voxelNeighborship}. These definitions define how pairs of touching voxels such as shown in Figure~\ref{fig:touchingVoxels_1} and ~\ref{fig:touchingVoxels_2} are treated,  whether connected and forming one surface or disconnected and forming two surfaces. That is, the number of entities in~\eqref{eq:eulerVol} and~\eqref{eq:eulerSurf} depends on the chosen definition of the neighbourship. For a voxel~$v$ its 6-neighbours consist of all the cells which share a face with~$v$ and its 26-neighbours consist of all the cells which share a common vertex with~$v$. The voxels in the  6-neighbourhood of~$v$ are denoted with~$\set N_{6}(v)$ and the ones in the 26-neighbourhood with~$\set N_{26}(v)$. In our implementation, two solid voxels are considered to be neighbours when they are 26-neighbours.  In contrast,  two void voxels are considered to be neighbours when they are 6-neighbours. As pointed out in Kong et al.~\cite{kong1989digital}, it is essential not to use the same neighbourship definitions for both the solid and void voxels. The use of one single neighbourhood definition leads to ambiguities, for instance, in the planar case to the violation of the Jordan curve theorem. The Jordan curve theorem states that a simple closed curve divides the plane into an interior and an exterior region.  

\begin{figure}[bt]
\centering
\subfloat[][$n_0^{(q)} = 1$] {
	\includegraphics[scale=0.5]{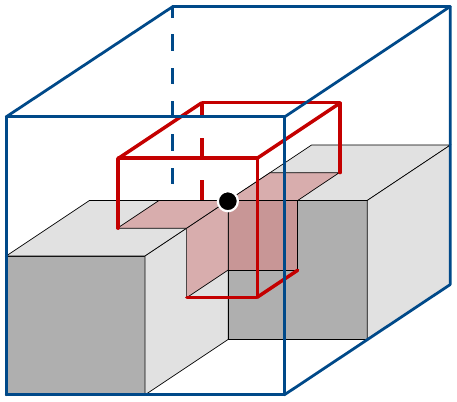} }
\hspace{0.075\textwidth}
\subfloat[][$\frac{n_1^{(q)}}{2} = \frac{5}{2}$] {
	\includegraphics[scale=0.5]{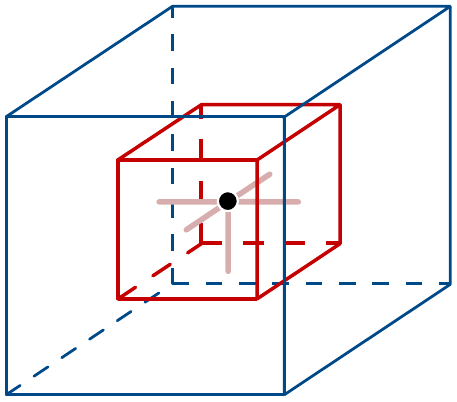} }
\hspace{0.075\textwidth}
\subfloat[][$\frac{n_2^{(q)}}{4} = \frac{6}{4}$] {
	\includegraphics[scale=0.5]{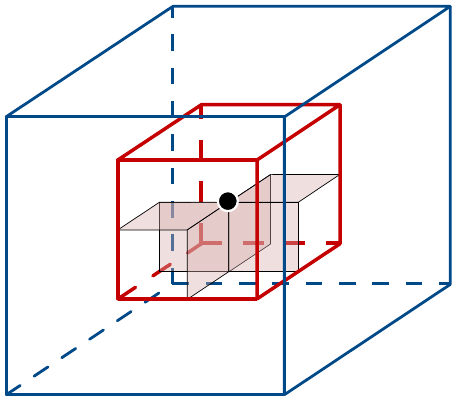}  }
\hspace{0.075\textwidth} 
\subfloat[][$\frac{n_3^{(q)}}{8} = \frac{2}{8}$] { 
	\includegraphics[scale=0.5]{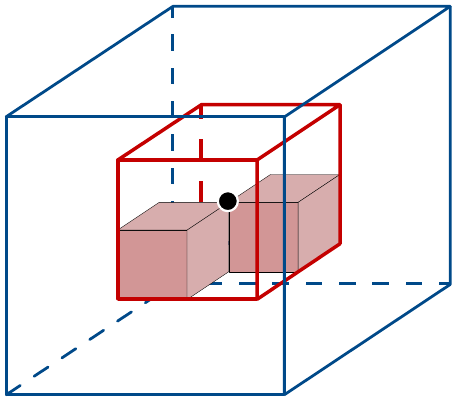}}
\caption{Contribution~$\chi(\set O(q)) = 1 - \frac{5}{2} + \frac{6}{4} - \frac{2}{8} = - \frac{1}{4}$ of the octant~$\set O(q)$ to the Euler characteristic. Example of an octant with two voxels.} \label{fig:eulerContrib}
\end{figure}

The Euler characteristic~\eqref{eq:eulerVol} and \eqref{eq:eulerSurf} for the entire voxel model can be determined by looping over all vertices in the grid and only considering in each step the voxels attached to one vertex. The  2$\times$2$\times$2 voxels attached to a vertex~$q$ are defined as its octant~$\set O(q)$.\footnote{Note that it is also possible to associate the octants with the voxels instead of the vertices in the grid. Both viewpoints are  equivalent~\cite{lee1994building}.}   There are as many octants as vertices in the volume grid (assuming that the voxel domain is padded with empty ghost voxels).  The octants are overlapping and the edges, faces and voxels in the hexahedral mesh~$\set M_s$ appear in several octants. Hence, in computing the Euler characteristic~\eqref{eq:eulerVol} by iterating over the vertices the contribution of each octant has to be suitably weighted such that  
\begin{equation}  \label{eq:eulerSum}
	\chi (\set M_s) = \sum_q \chi (\set O(q)) = \sum_{q} \left  (1 -\frac{n_1^{(q)}}{2} + \frac{n_2^{(q)}} {4} -   \frac{n_3^{(q)}} {8} \right )  \, ,
\end{equation}
where~$n_1^{(q)}$ is the number of  edges and~$n_2^{(q)}$ is the number of faces neighbouring to the~$n_3^{(q)}$ solid voxels in the mesh~$\set M(\set O(q)) \setminus \partial \set M(\set O(q))$, see Figure~\ref{fig:eulerContrib}. The contribution of an octant $\chi(\set O(q))$ with different solid-empty voxel states can be precomputed and stored in a look-up table. The eight voxels in an octant have $2^8=256$ possible solid-empty states and considering symmetries this reduces to only 22 distinct cases. Tables with contributions of octants to the Euler characteristic can be found, e.g., in~\cite{lee1994building,homann2007,post2016fast}
%

%
\subsection{Skeletonisation by successive voxel removal \label{sec:skeletonisation}}
\begin{figure}
	\centering
	\includegraphics[scale=0.38, angle=90]{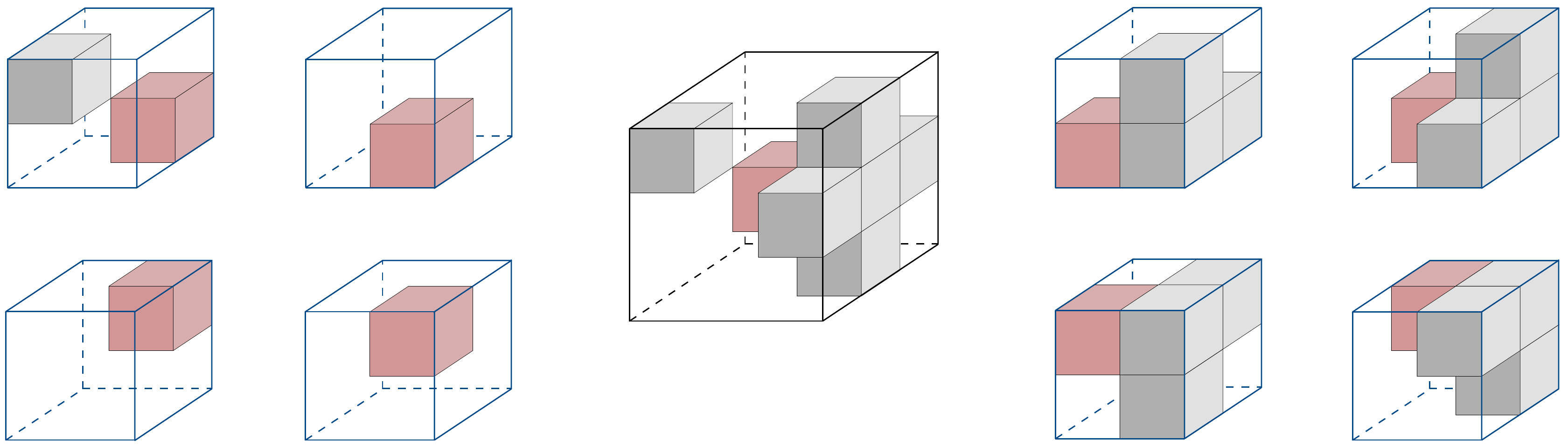}
	\caption{  26-neighbours of a voxel (red) and the eight overlapping octants used for determining the change in the Euler characteristic before and after the voxel is deleted. The contributions of the eight octants with the voxel present are -$\frac{3}{4}$, $\frac{1}{8}$, -$\frac{1}{4}$, -$\frac{1}{4}$, $\frac{1}{8}$, $\frac{1}{8}$, $-\frac{1}{4}$  and $-\frac{1}{4}$ with the voxel removed are  $\frac{1}{8}$, $0$, $-\frac{1}{8}$, $-\frac{1}{8}$, $0$, $0$, $-\frac{1}{8}$ and $-\frac{1}{8}$ (clockwise), see~\cite{lee1994building}. Hence, the voxel cannot be removed because this would lead to a change in the Euler characteristic by $\Delta \chi = 1$.
	\label{fig:changeInEuler}}
\end{figure}

%
The Euler characteristic~$\chi (\set M_s) $ is critical in determining the voxels at the boundary of the solid mesh~$\set M_s$  that can be deleted without changing the topology of the voxel model.  A {\em solid border voxel} is defined as a solid voxel with at least one void voxel amongst its 6-neighbours. There is a large amount of research on voxels, also referred to as {\em simple points},  which can be removed without changing topology~\cite{kong1989digital}. As evident from~\eqref{eq:eulerGlobal} simply conserving the Euler characteristic~$\chi(\set M_s)$ of the entire or a portion of the mesh does not ensure that the topology is not changed, i.e. the number of objects~$o(\set M_s)$, handles~$h(\set M_s)$ and cavities~$c(\set M_s)$ all remain the same.  For a solid border voxel~$v$ to be classified as simple its removal must not change, in addition to the Euler characteristic,  the number of objects and handles for both~$\set M_s$ and~$\set M_e$~\cite{kong1989digital}.  According to Lee et al.~\cite{lee1994building} these conditions can be checked by examining the state of the  voxels in eight octants overlapping voxel~$v$, or the voxels in $v$'s  26-neighbourhood~$\set N_{26}(v)$. That is, a border voxel~$v$ is a simple point if and only if  
\begin{subequations} \label{eq:changeCrit}
\begin{align}
	\Delta \chi  (\set M \left ( \set N_{26}(v))  \right )  &= 0 \quad  \text{ and }  \label{eq:changeChi}  \\
	\Delta h \left (\set M(\set N_{26}(v))  \right ) &= 0  \quad \text{ or }  \\ 
	\Delta o \left ( \set M(\set N_{26}(v))  \right ) &= 0  \label{eq:changeObj} \, ,
\end{align} \label{eq:simpleCheck}
\end{subequations}
where~$\Delta$ denotes the change of the respective quantity with and without voxel~$v$ present. It is straightforward to determine the Euler characteristic~\eqref{eq:changeChi} and it is relatively easy to determine the number of objects~\eqref{eq:changeObj} in~$\set N_{26}(v)$.  The change of the Euler characteristic is determined by summing the tabulated contributions of the eight octants belonging to the eight corners of the voxel~$v$ from~\cite{lee1994building}, see ~\eqref{eq:eulerSum} and Figure~\ref{fig:changeInEuler}.  To determine the number of objects~\eqref{eq:changeObj}, we generate an undirected graph with the solid voxels as nodes and introducing edges between voxels that are 26-neighbours. Subsequently, the number of connected components is determined with a depth first search (DFS) algorithm \cite{even2011graph}. 

The skeletonisation proceeds by removing in each step one layer of border voxels which are simple points according to~\eqref{eq:simpleCheck}, see Algorithm~\ref{alg:skeletonisation} in~\ref{sec:appendixAlg}. The removed solid voxels are reclassified as void.  One removal step is split into six sub-steps and in each sub-step only the voxels approaching from one of the six grid directions are removed. The geometry of the obtained skeleton depends somewhat on the sequencing of these sub-steps, which is not critical for our purpose. It is straightforward to tag some voxels as non-removable irrespective of~\eqref{eq:simpleCheck}. In skeletonising topology optimisation results, for instance, the voxels at Dirichlet and non-zero  Neumann boundaries are tagged as non-removable. Also,  end points with only one voxel in their 26-neighbourhood have to be tagged as non-removable, in order to avoid a complete elimination of voxel chains. The skeletonisation terminates when none of the remaining voxels is removable without violating~\eqref{eq:simpleCheck}. The final result is a curve skeleton consisting of  a network of  one-voxel-wide voxel chains connected by joint voxels. We refer to this curve skeleton as the {\em voxel chain skeleton}.

%
\subsection{Illustrative example and timing}
\begin{figure}[t]
	\centering
	\subfloat[][CAD model]
	{
		\includegraphics[width=0.35\textwidth]{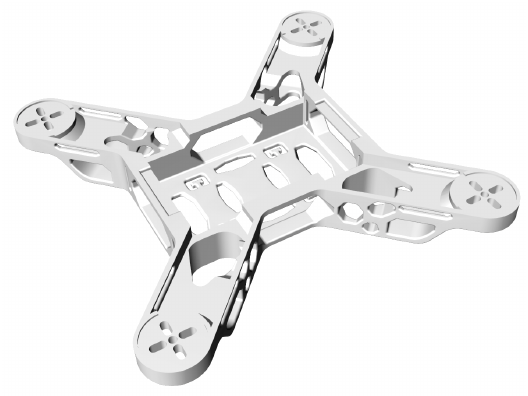}
		\label{fig:quadCopter_model}
	}\hspace{0.15\textwidth}
	\subfloat[][STL mesh]
	{
		\includegraphics[width=0.35\textwidth]{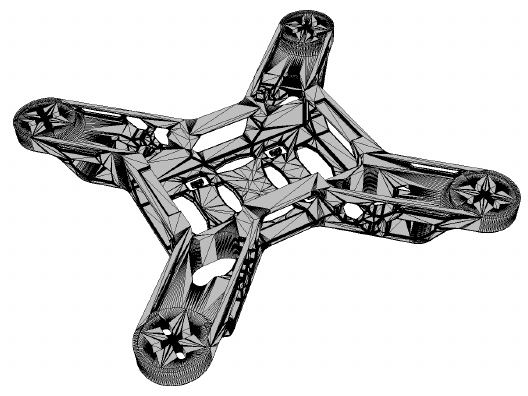}
		\label{fig:quadCopter_stl}
	}\\
	\subfloat[][Voxel model]
	{
		\includegraphics[width=0.35\textwidth]{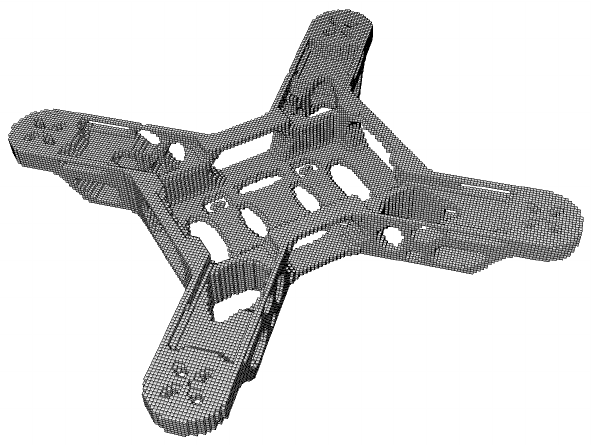}
		\label{fig:quadCopter_voxel}
	}\hspace{0.15\textwidth}
	\subfloat[][Voxel chain skeleton]
	{
		\includegraphics[width=0.35\textwidth]{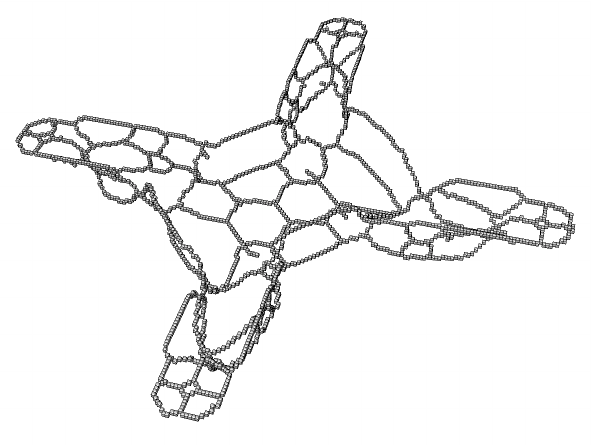}
		\label{fig:quadCopter_skeleton}
	}
	\caption{Skeletonisation of a quadcopter frame designed in a CAD system. The voxel model in (c) is obtained from the STL mesh in (b) by computing its implicit, or level set, representation on a uniform voxel grid and thresholding. The voxel chain skeleton is determined from the voxel model in (c) with the introduced skeletonisation algorithm. For further details see under Grid 3 in Table~\ref{tab:thinningPerformance}. 	
}
	\label{fig:quadCopter}
\end{figure}

%
We consider the skeletonisation of the quadcopter frame shown in Figure~\ref{fig:quadCopter_model} obtained from GrabCAD. The frame has a non-trivial topology and has been designed in a CAD system. To generate a voxel model the implicit, or level set, representation of the frame on a uniform voxel grid is computed first.  This is achieved by determining the distance of each voxel in the grid to the frame surface with the open source openVDB library~\cite{museth2013openvdb}.  In doing so, the frame geometry is approximated with the STL mesh exported from the CAD system depicted in Figure~\ref{fig:quadCopter_stl}. After thresholding the voxel grid the voxel model of the frame in Figure~\ref{fig:quadCopter_voxel} is obtained. Its skeletonisation with the introduced digital topology algorithm yields the voxel chain skeleton in Figure~\ref{fig:quadCopter_skeleton}. 
 It can be visually confirmed that the skeleton appears to have the same topology as the voxel model. 

To investigate the efficiency and scaling of our C++ implementation of  the introduced skeletonisation algorithm, we consider three different voxel grids for computing the implicit representation and skeletonisation. The size of the three grids, the number of voxels in the different models and the runtimes are given in Table~\ref{tab:thinningPerformance}. The STL mesh has in all cases 1086791 triangles.  The number of solid voxels in the voxel model and the number of removal steps increase with decreasing voxel size because more and more of them cover the frame. The number of skeleton voxels increases because the smaller voxels can capture the topology of the frame better.  The reported extremely short runtimes include only skeletonisation (no implicitisation) and confirm the efficiency of the skeletonisation algorithm. Moreover, notice that the average time per removal step is approximately linear with respect to the number of the voxels in the voxel grid. The experiments were performed on a Macbook Pro with an Intel Core CPU i7-4750HQ  $@$ 2.0GHz and 16GB~RAM.
\begin{table}[b]
	\centering
	\begin{tabular}{c c c c c c c}
		\hline
		& \begin{tabular}{c}
			Voxel grid \\
			\# voxels
		\end{tabular} & 	
		\begin{tabular}{c}
			Voxel model \\
			\# voxels
		\end{tabular} & 
		\begin{tabular}{c}
			Skeleton \\
			\# voxels	
		\end{tabular} &
		\begin{tabular}{c}
				\\
			removal  
			steps
		\end{tabular} &
		\begin{tabular}{c}
			\\ 
			time per step  
		\end{tabular} & 		
		\begin{tabular}{c} 
			\\ 
			total time
		\end{tabular} \\
		\hline
		Grid 1 & $139\times16\times139$ & 18222 & 1278 & 4 & $\phantom{0}4.43$ s & $17.73$ s \\
		Grid 2 & $172\times19\times172$ & 38076 & 1677 & 5 & $\phantom{0}9.33$ s & $46.65$ s \\
		Grid 3 & $197\times21\times197$ & 53981 & 1894 & 6 & $13.17$ s & $78.99$ s \\ \hline
	\end{tabular}
	\caption{Efficiency and scaling of the skeletonisation for the quadcopter frame on three different voxel grids.}
	\label{tab:thinningPerformance}
\end{table}

%
\section{Model extraction and conversion \label{sec:extractionAndCovnversion}}
%
The voxel chain skeleton is used to define a structural frame model.  The skeleton provides both the  connectivity and the geometry of  the frame.  Although it is feasible to obtain also the member cross-sections from the voxel model, in our implementation cross sections are obtained from size optimisation of the frame, as will be discussed in Section~\ref{sec:overall}.  In  converting the voxel chain skeleton to a frame we express it as a weighted undirected graph and make use of its incidence matrix. The frame model provides a very compact representation of the optimised structure, which can be used to generate a parametric solid model using the scripting interface or API (application programming interface) of a  CAD system. 
%
\subsection{Graph model \label{sec:graphModel}}
\begin{figure}[]
		\centering
		\subfloat[Voxel model]
		{
			\includegraphics[scale=2.]{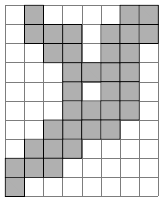}
			\label{fig:2dSkeletonisedA}
		}
\hspace{0.125\textwidth}
		\subfloat[Voxel chain skeleton]
		{
			\includegraphics[scale=2.]{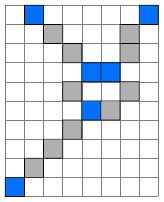}
			\label{fig:2dSkeletonisedB}
		}
		\caption{Two dimensional illustrative example for the skeletonisation of a voxel model. In the voxel chain skeleton the grey voxels with two neighbours denote the regular voxels and the blue voxels with other than two neighbours are the joint voxels.  \label{fig:2dSkeletonised}}
\end{figure}

%
The voxel chain obtained from skeletonisation contains two different types of voxels, see Figure~\ref{fig:2dSkeletonised}.  We refer to the voxels  with only two voxels in their 26-neighbourhood as {\em regular voxels}. The remaining voxels are the {\em joint voxels}, which have other than two voxels in their 26-neighbourhood.  As illustrated in~Figure~\ref{fig:2dSkeletonisedB}, it is not expedient to deduce from every joint voxel a joint for the frame model because this will lead to acutely short members and impractical joint designs. 

\begin{figure}[]
	\centering
	\begin{minipage}{\textwidth}
		\centering
		\subfloat[Graph with the edge $e_3$ to be collapsed]
		{
			\includegraphics[scale=0.95]{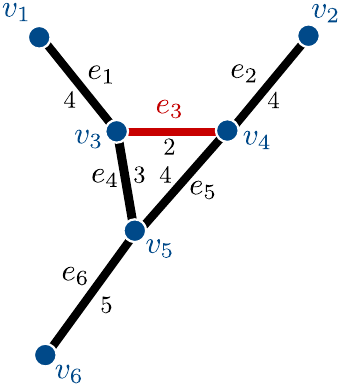}
			\label{fig:graphModel}
		}
		\hspace{0.12\textwidth}
		\subfloat[Graph after edge collapse with duplicate edge]
		{
			\includegraphics[scale=0.95]{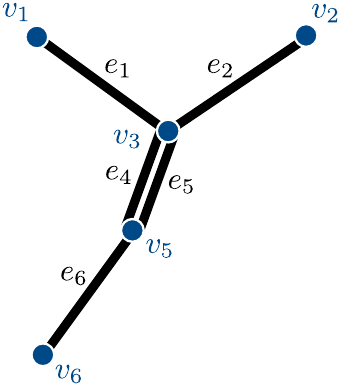}
			\label{fig:duplicateEdges}
		}
		\hspace{0.12\textwidth}
		\subfloat[Final graph after  the removal of duplicate edge]
		{
			\includegraphics[scale=0.95]{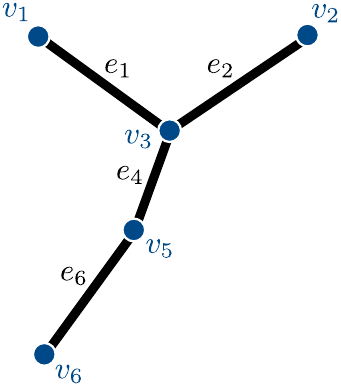}
		}
	\end{minipage}
	\begin{minipage}{\textwidth}
		\centering
		\subfloat[Incidence matrix]
		{
			\includegraphics[scale=1]{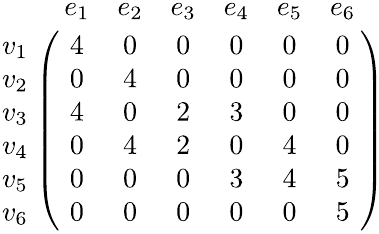}
			\label{fig:incidenceMatrix}
		}
		\hspace{0.12\textwidth}
		\subfloat[Edge collapse]
		{
			\includegraphics[scale=1]{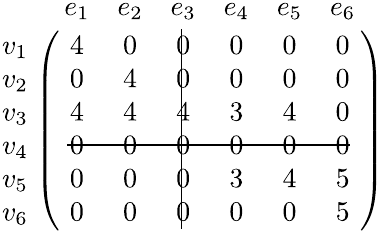}
		}
		\hspace{0.12\textwidth}
		\subfloat[Duplicate edge removal]
		{
			\includegraphics[scale=1]{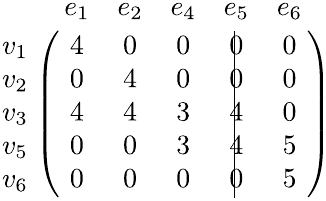}
			\label{fig:dupliateMatrix}
		}
	\end{minipage}
	\caption{Merging of graph nodes connected by short edges using edge collapse. The edge $e_3$ is collapsed by merging the row~$v_3$ with row~$v_4$ and removing the column $e_3$ in the incidence matrix. The duplicate edge $e_5$ after the collapse is merged with~$e_4$ by averaging their weights and deleting the column $e_5$.}
	\label{fig:mergeProcess}
\end{figure}

To robustly merge joint voxels in close proximity to a single joint we resort to a weighted undirected graph representation of the voxel chain skeleton. The nodes of the graph are the joint voxels, and the edges are the connections between the joint voxels. Each of the nodes has an associated coordinate vector which is initially equal to the centroid of the corresponding joint voxel. The edge weights are proportional to the number of voxels between the two joints attached to an edge. The graph model is generated with the Algorithm~\ref{alg:extractGraph} given in \ref{sec:appendixAlg}.  In a first step, all the solid voxels in the voxel model are categorised either as regular or joint. Subsequently, starting from the joint voxels we determine the voxel chains and their lengths by marching along the $\mathcal{N}_{26}(v)$  neighbours until a joint voxel is reached.  The obtained graph model is stored in the form of an incidence matrix. In Figure~\ref{fig:graphModel}  and~\ref{fig:incidenceMatrix} the graph model and its incidence matrix for the voxel chain skeleton in Figure~\ref{fig:2dSkeletonisedB} are given. The rows of the incidence matrix correspond to the nodes of the graph and the columns to its edges. Each edge has two identical non-zero entries corresponding to the length of the edge in voxel units.

The joint voxels in close proximity of each other are connected by very short edges. The corresponding graph nodes are merged by successively collapsing the edges of the graph. The edge collapse is implemented with the help of the incidence matrix. Figure~\ref{fig:mergeProcess} illustrates how the incidence matrix evolves during the collapse of the edge~$e_3$  and the merging of its attached nodes~$v_3$  and~$v_4$. The coordinate vector associated to the node~$v_3$ is the average of the node coordinate vectors before merging. After collapsing, the duplicate edge~$e_5$ is detected and merged with~$e_4$. The weight of~$e_4$ is updated to be the closest integer to the mean of the weights of $e_4$ and $e_5$. 

\begin{figure}[]
	\centering
	\subfloat[Node $v_1$ and edge~$e_1$ to be pruned]
	{
		\includegraphics[scale=0.95]{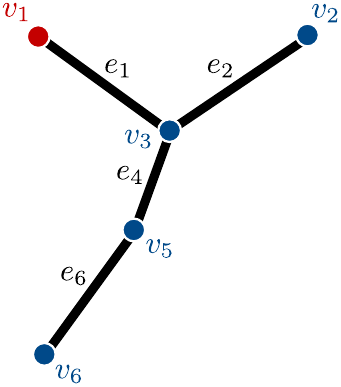}
	}
	\hspace{0.12\textwidth}
	\subfloat[Graph after pruning]
	{
		\includegraphics[scale=0.95]{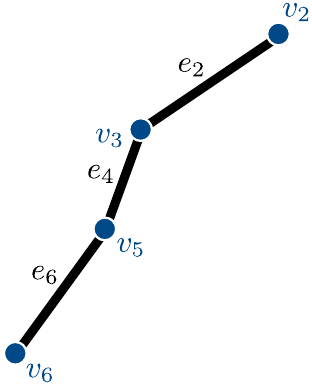}
		\label{fig:prunedModel}
	}
	\hspace{0.12\textwidth}
	\subfloat[Incidence matrix]
	{
		\includegraphics[scale=1]{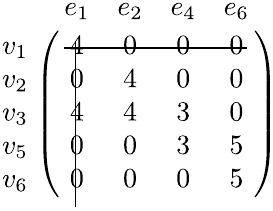}
		\label{fig:pruningMatrix}
	}
	\caption{Pruning of  branches which lead to a structural frame member with zero stress. The node $v_1$ and the  attached edge $e_1$  may be deleted if~$v_1$ is not on a Dirichlet or non-homogeneous Neumann boundary. The node and edge are deleted by removing the corresponding row and column of the incidence matrix.}
	\label{fig:pruningProcess}
\end{figure}

The voxel chain model usually contains branches which lead to structural frame members with zero stress. 
These are the chains directly connected to a zero Neumann boundary. 
Their pruning is again implemented with the incidence matrix. As illustrated in Figure~\ref{fig:pruningProcess}, after identifying from the incidence matrix the nodes with only one attached edge, nodes and edges are deleted by simply  removing the corresponding rows and columns.

\subsection{Structural frame and  CAD models \label{sec:CADimport}}
%
The graph model supplemented by the joint coordinates and member cross-sections  provides sufficient information to generate a compact structural frame model. We assume that there is a one-to-one correspondence between nodes and edges of the graph model and joints and members of the frame model. Besides, it is assumed that members are only subjected to end forces and moments, i.e. no distributed loads,  so that they can be straight  and have uniform cross-sections along their lengths. Note that these assumptions are fully justified if there is no distributed loading in  topology optimisation. Furthermore, for simplicity, we assume in the following that all cross-sections are circular.  These restrictions can be mostly relaxed if necessary.

\begin{figure}[]
	\centering
	\includegraphics[width=0.7\textwidth]{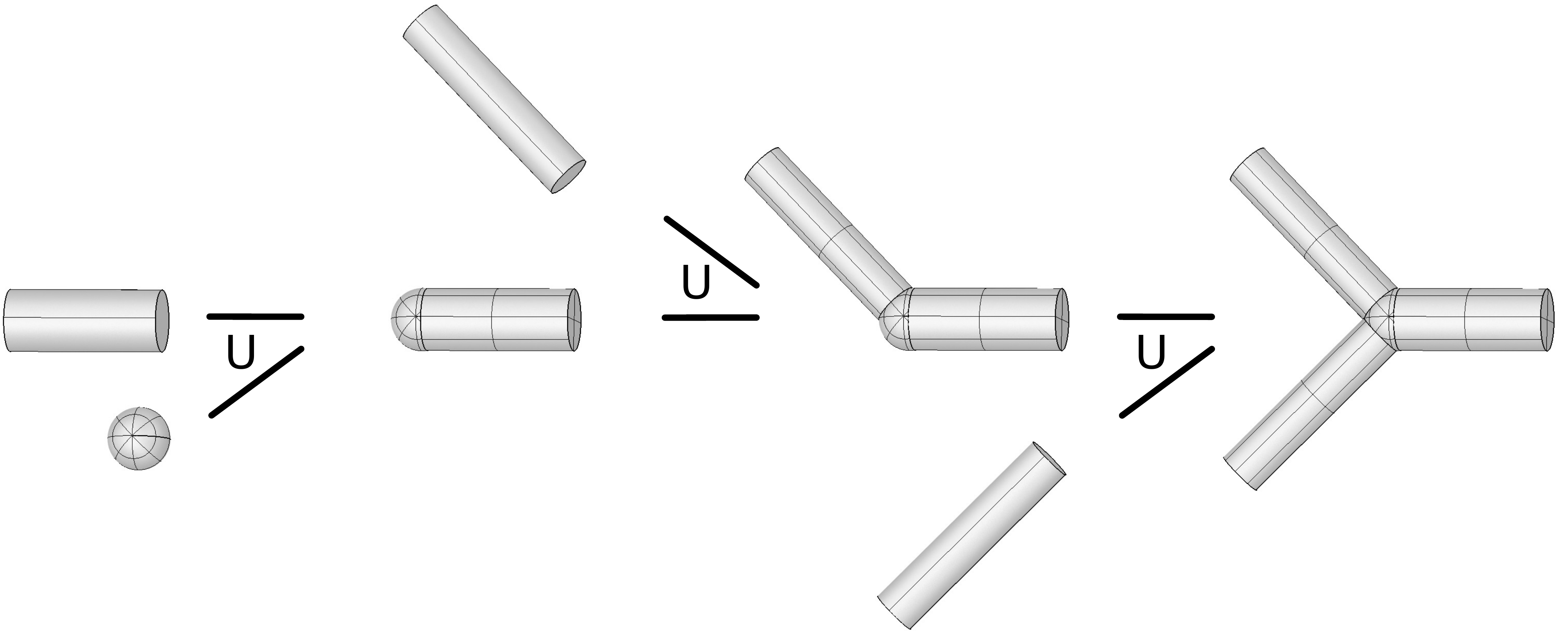}
	\caption{Binary CSG tree of a CAD model generated by successively combining cylinders with spheres using boolean union.}
	\label{fig:csgTree}
\end{figure}

It is straightforward to create a parametric solid CAD model from the structural frame model. To this end, we use either the commercial CAD system Rhinoceros (Rhino)  or the open source FreeCAD. However, the proposed approach can be realised in any CAD system which provides a  scripting interface or API (application programming interface).  In the solid model, the members are represented by cylinders and the joints by spheres, see Figure~\ref{fig:csgTree}. They are combined with boolean operations. The radius of the sphere  at a joint is chosen slightly larger than the largest attached member radius. In this paper, we use a factor of~$1.05$.  A binary CSG tree of the frame model is generated by starting with one of the members and by successively adding spheres and cylinders to the model.  
Different variations of this approach can be devised and implemented.  For instance, it suggests itself to connect two of the most stressed members continuously across a joint or to add fillets to reduce stress concentrations. See also~\cite{smith2016application, Arora:michell:scf:2019} in which a somewhat similar approach was used for generating CAD models for optimised truss structures.  After the model is generated it can be edited and postprocessed  in the CAD system and be exported, for instance, in STL format for machining or IGES or STEP formats for mesh generation software. 

%
\section{Overall optimisation and conversion process \label{sec:overall}}
%
In this Section, we review the sequence of steps from the definition of  the topology optimisation problem to obtaining the structurally-sound parametric CAD geometry. We provide additional details for each of the steps and  focus on the interplay between the different steps. Each step corresponds to one Section in this paper. 

\paragraph{Step 1: Topology optimisation}  We assume that the finite element mesh for the optimisation problem~\eqref{eq:topOpt} is a structured hexahedral grid. In most optimisation problems, the design domain is a parallelepiped which can be discretised with a structured grid.  Other more complex design domains can be considered by computing their implicit, or level set, representation and embedding them in a structured hexahedral grid, see  the quadcopter frame example in Figure~\ref{fig:quadCopter}. From the outset, the voxels outside the design domain are chosen as void by choosing their Young's moduli with~$E_{\text{min}}$.

\paragraph{Step 2: Skeletonisation} The input to the homotopic skeletonisation algorithm is a binary image defined on a hexahedral structured grid. The binary image is obtained by thresholding the topology optimised geometry. The threshold value~$\eta$ is chosen such that the prescribed material volume constraint~\eqref{eq:topOptC} in topology optimisation is retained. The output of the skeletonisation is a curve skeleton in the form of a voxel chain with the same topology as the topology optimised geometry.

\paragraph{Step 3: Structural  frame model generation} The topology and joint coordinates of the frame model are extracted from the voxel chain model. It is assumed that all the members are straight, have a circular cross-section with the same diameter and their total volume is equal to the prescribed material volume~$V_f \overline{V}$  in topology optimisation.  It is possible to obtain frame models with more complex member geometries and non-uniform cross-sections.  This appears to be, however, usually not desired from an ease and cost of manufacturability viewpoint. 

\paragraph{Step 4: Size and layout optimisation} The structural frame model extracted from the skeleton is usually suboptimal. That is, the compliance of the structural frame is larger than the compliance of the topology optimised geometry. 
To recover the optimality of the frame structure, we apply several steps of sequential size and layout optimisation~\eqref{eq:frameOpt}.  In size optimisation the diameters~$d_i$ of the circular member cross-sections and in layout optimisation the coordinates~$\vec x_i$ of the joints are updated. Both optimisation problems  are solved with the SQP (sequential quadratic programming) method~\cite{nloptPackage}. Throughout optimisation the volume of the frame is constrained to be equal to the prescribed material volume in topology optimisation~$V_f \overline{V}$.  It is straightforward to consider additional constraints pertaining to the member cross-sections, the positions of the joints, or positions and orientations of the members.  The first step is always size optimisation, which is followed by as many as necessary alternating layout and size optimisation steps until the compliance cost function is converged. If the length of any member reduces to zero during shape optimisation, the member is removed, its end nodes are merged, and the iteration continues.

\paragraph{Step 5: CAD model generation} A compact CAD model of the structural frame can be generated essentially in any parametric CAD system using a fully automated process. The members are represented by cylinders and the joints by spheres which are combined by boolean operations. The underlying binary CSG tree representation makes it easy to edit further and to refine  the optimised design.

%
\section{Examples \label{sec:examples}}
%
We consider three examples of increasing complexity to demonstrate the application of the proposed approach. The minimised cost function  is the compliance~$J(\hat{\vec \rho})$ in topology optimisation and~$J(\vec d, \, \vec x)$ in size and layout optimisation, where~$\hat{\vec \rho}$ are the filtered element densities, $\vec d$ are the member diameters, and $\vec x$ are the joint coordinates.  In all examples, the Young's modulus and Poisson's ratio of the solid material are $\overline E=2.1 \cdot 10^5$ and $\nu=0.3$. In topology optimisation the penalisation factor is chosen as $p = 3$ and the minimum Young's modulus as $E_{min}=10^{-9}$. During layout optimisation, a member shorter than $1/20$ of the total length of its immediate neighbouring members is considered as too short, and its two end nodes are merged.  
%
\subsection{Cantilever \label{sec:cantilever}}
\begin{figure}[]
	\centering
	\includegraphics[width=0.475\textwidth]{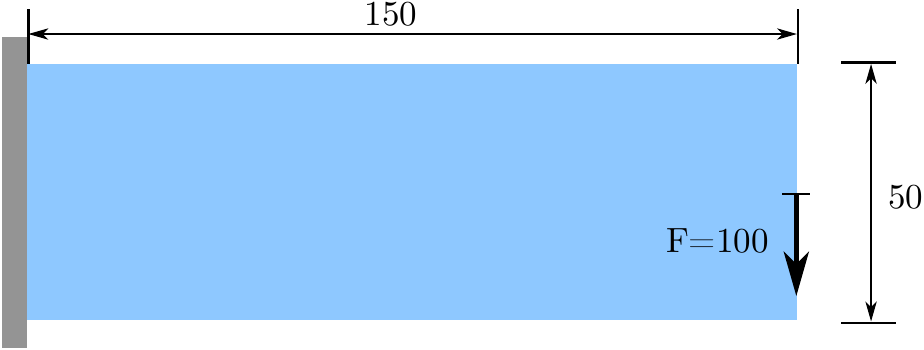}
	\caption{Geometry, boundary conditions and loading of the cantilever.}
	\label{fig:cantilever_system}
\end{figure}

\begin{figure}[]
	\vspace{1.em}
	\centering
	\subfloat[][Thresholded geometry with $\eta=0.3$ \label{fig:cantilever_top} ] {	
		\includegraphics[width=0.375 \textwidth]{examples/cantilever/cantilever_top.pdf}
	}
	\hspace{0.125 \textwidth}
	\subfloat[][ Voxel chain skeleton \label{fig:cantilever_skeleton}  ] {	
		\includegraphics[width=0.375\textwidth]{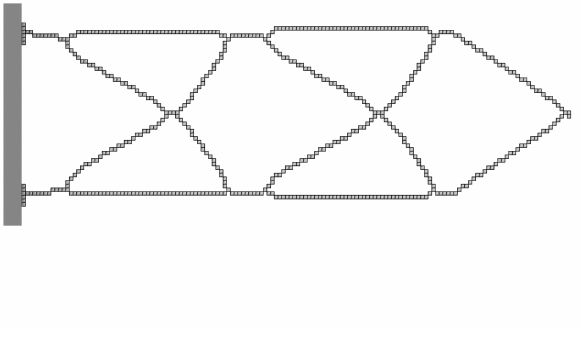}
	}
	\caption{Topology optimised and skeletonised cantilever.}
\end{figure}

%
Cantilevered plates as shown in Figure~\ref{fig:cantilever_system} are one of the most widely studied benchmark examples in topology optimisation, see e.g.~\cite{bendsoe2003topology}. The length, height and width of the chosen design domain are $150 \times 50 \times 4$. The left face of the domain is clamped while all the other faces are free, and a distributed force with a total value of $F = 100$ is applied along the centre of the right face. The finite element discretisation consists of $150 \times 50 \times 4$ linear hexahedral elements. The maximum volume fraction of the optimised structure is prescribed to be $V_f = 0.3$ and the density filter radius is chosen as $R=3$. Figure~\ref{fig:cantilever_top} depicts the optimised cantilever with only the voxels above a relative density $\eta=0.3$. The compliance of this voxel model is $J(\hat{\vec \rho})=3.40900$, which has been determined after topology optimisation by resetting $p=1$ in~\eqref{eq:younPenalised}. 

\begin{figure}[]
	\centering 
	\subfloat[][Size optimised (initial) \label{fig:cantilever_frame}] {	
		\includegraphics[width=0.375\textwidth]{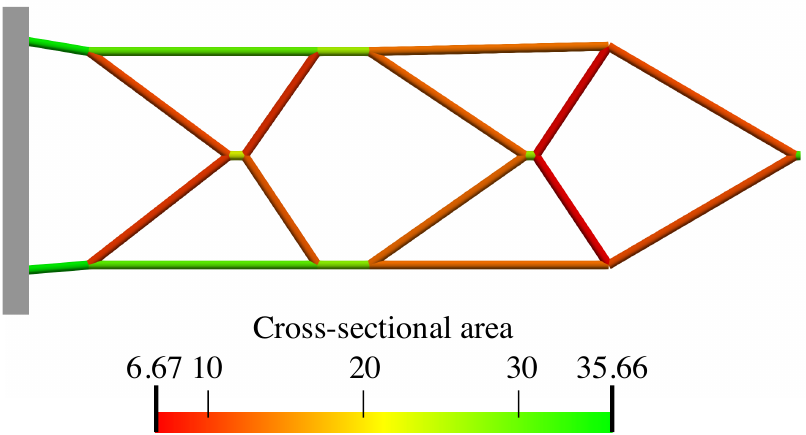}
		}
	\hspace{0.125\textwidth}
	\subfloat[][Size and layout optimised (final) \label{fig:cantilever_result}] {
		\includegraphics[width=0.375\textwidth]{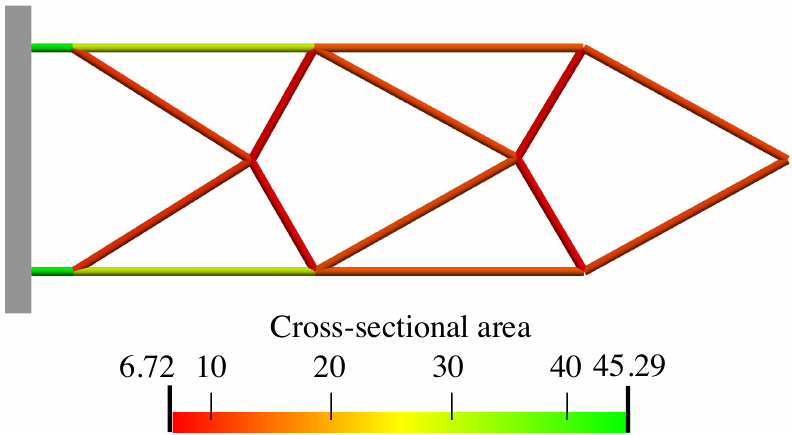}
	}
	\caption{Size and layout optimised cantilever frame.}
\end{figure}

As a first step in obtaining the structural frame model the homotopic skeletonisation algorithm is employed. The skeletonisation yields after $5\times6$ voxel removal steps the voxel chain skeleton shown in Figure~\ref{fig:cantilever_skeleton}. As visually apparent, the voxel and voxel chain models have the same topology.  The skeleton is converted into the structure depicted in Figure~\ref{fig:cantilever_frame} by first identifying the 16 joints and then connecting them with 21 members.  As evident from Figure~\ref{fig:cantilever_frame} during the conversion to a frame model the optimality of the voxel model is compromised; note, for instance, the non-straight top and bottom members. Optimality is recovered by updating the joint positions, i.e. layout optimisation, and the cross-sectional areas, i.e. size optimisation, see Figure~\ref{fig:cantilever_result}. Initially, all members are assumed to have the same diameter of ~$d=4.547$, giving the prescribed total volume of $V = 0.3 \overline V$. In determining the volume only the member cross-sections and member lengths between the joints are considered.

\begin{figure}[]
	\vspace{0.5em}
	\centering
	\hspace{0.5pt}
	\subfloat[][Topology optimisation \label{fig:cantilever_conv_top}]
	{
		\includegraphics[scale=0.5]{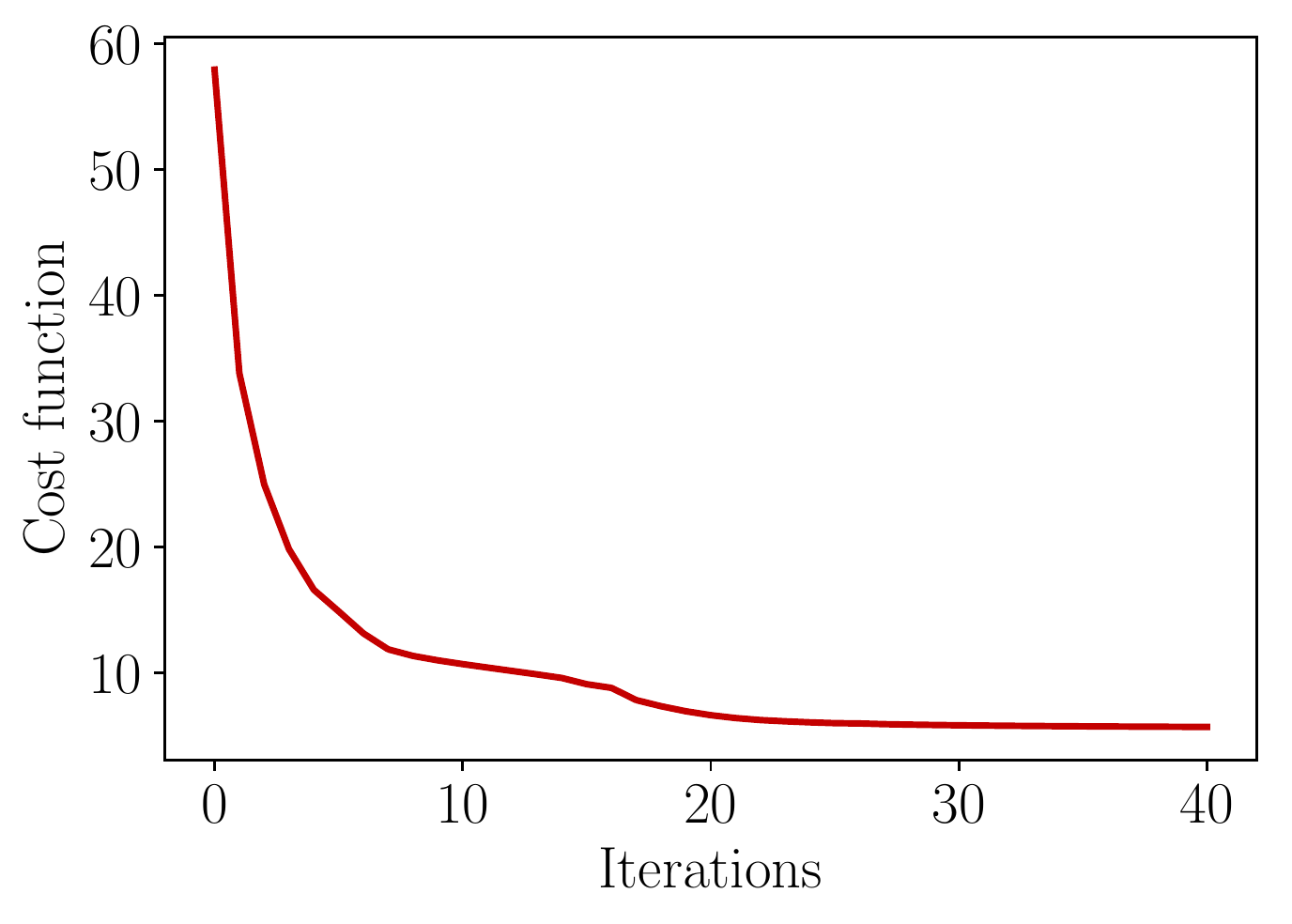}
	}	
	\hfill
	\subfloat[][Size (S) and layout (L) optimisation ]
	{
		\includegraphics[scale=0.5]{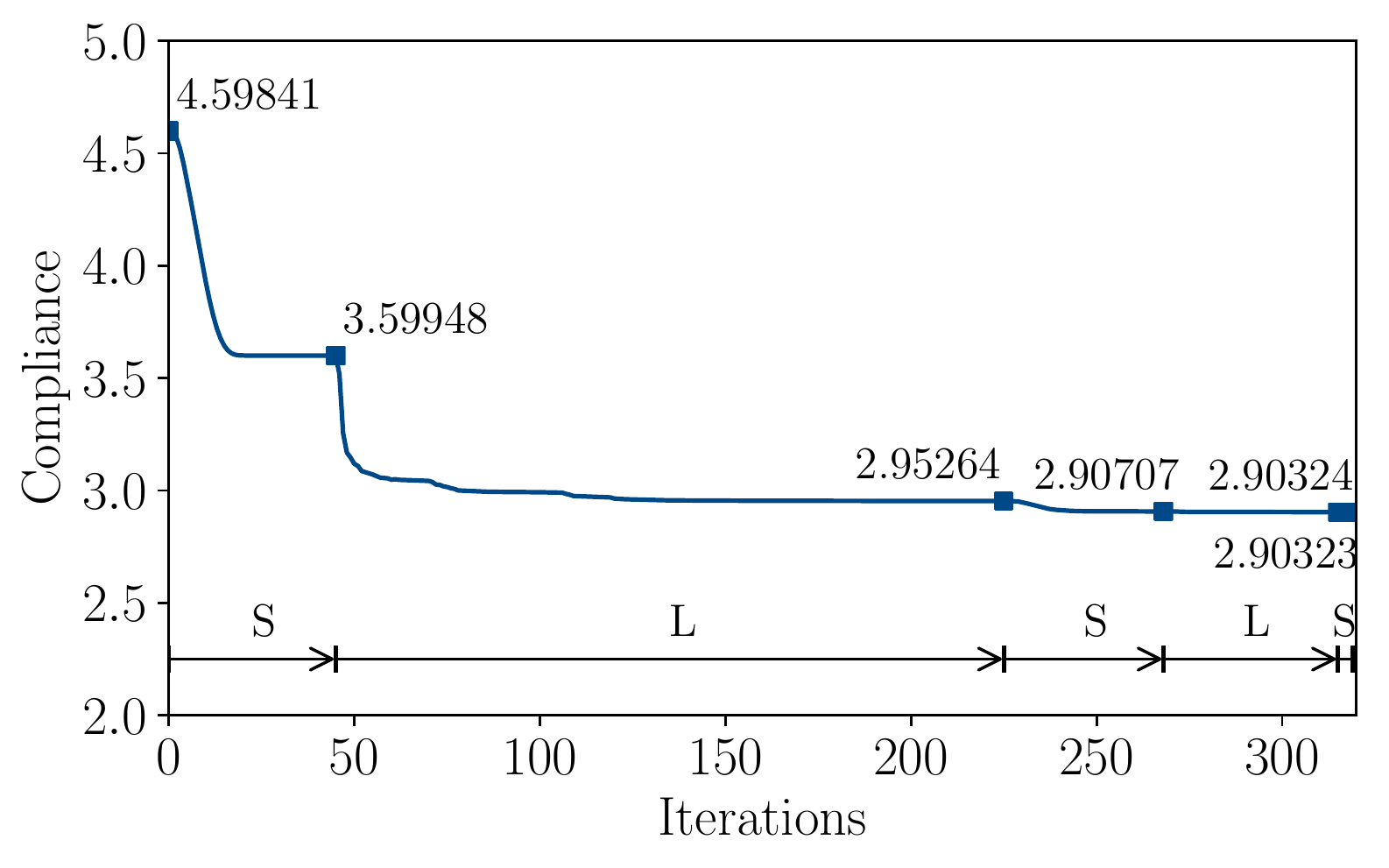}
		\label{fig:cantilever_conv_frame}
	}
	\caption{Convergence of the compliance during topology and sequential size and layout optimisation of the cantilever frame. Compliance of the topology optimised voxel model is $J(\hat{\vec \rho})=3.40900$. \label{fig:cantilever_conv}}
\end{figure}

\begin{figure}
	\vspace{0.5em}
	\centering
	\subfloat[][IGES model \label{fig:cantilever_iges}]
	{
		\includegraphics[width=0.4\textwidth]{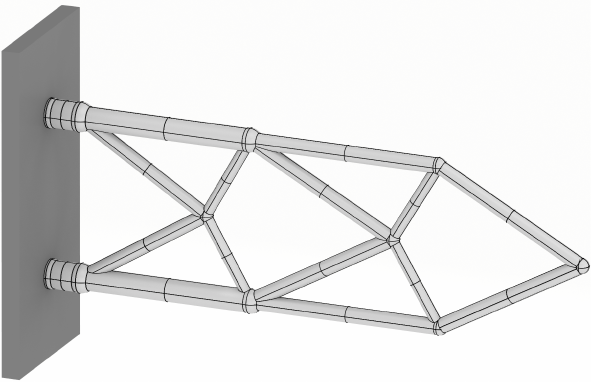}
	}
	\hspace{0.1\textwidth}
		\subfloat[][STL model \label{fig:cantilever_stl}]
	{
		\includegraphics[width=0.4\textwidth]{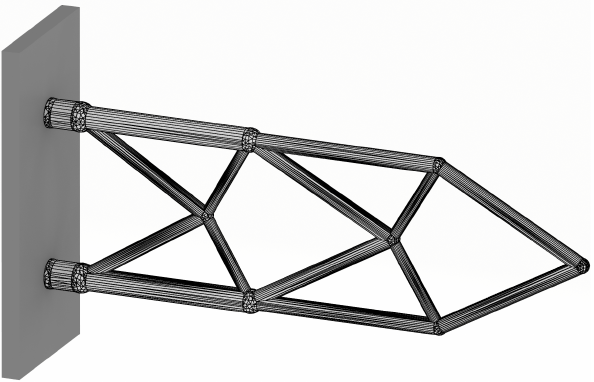}
	}
	\caption{Parametric CAD model of the cantilever. \label{fig:cantilever_cad}}
\end{figure}

According to Figure~\ref{fig:cantilever_conv_frame}, the frame consisting of beams with uniform diameter has a compliance \mbox{$J(\vec d, \vec x) = 4.59841$}, which is larger than the voxel model. The stretch and bending strain energies of the frame are 2.02202 and 0.247108, respectively. The minimum and maximum axial stresses are 0 and 20.9274. The average axial stress is 8.04859 and its standard deviation is 6.26102. After the first size optimisation step the compliance is reduced to \mbox{$J(\vec d, \, \vec x) = 3.59948$} and the subsequent layout optimisation step to \mbox{$J(\vec d, \, \vec x)=2.95264$}. Several more steps of size and layout optimisation do not lead to a significant reduction in the compliance. Notice that the obtained final compliance \mbox{$J(\vec d, \, \vec x) = 2.90323$}  is significantly lower than the compliance \mbox{$J(\hat{\vec \rho})= 3.40900$}  of the topology optimised voxel model.  The stretch and bending strain energies of the final frame are 1.42114 and 0.0190822, respectively. The minimum and maximum axial stresses are 7.00854 and 8.46234. The average axial stress is 8.09638 and its standard deviation is 0.44709. In the final optimised cantilever  in Figure~\ref{fig:cantilever_result} the axial stresses become more evenly distributed and the bending stresses are significantly reduced. The final optimised cantilever consists of 11 joints and 16 members.  Finally, in Figure~\ref{fig:cantilever_cad} the solid CAD model of the frame and a faceted triangular STL mesh exported from FreeCAD are shown. The lines in the CAD model show the layout of the trimmed NURBS patches.  

%
\subsection{Pipe bracket \label{sec:pipeHolder}}
\begin{figure}[]
	\centering
	\includegraphics[width=0.475\textwidth]{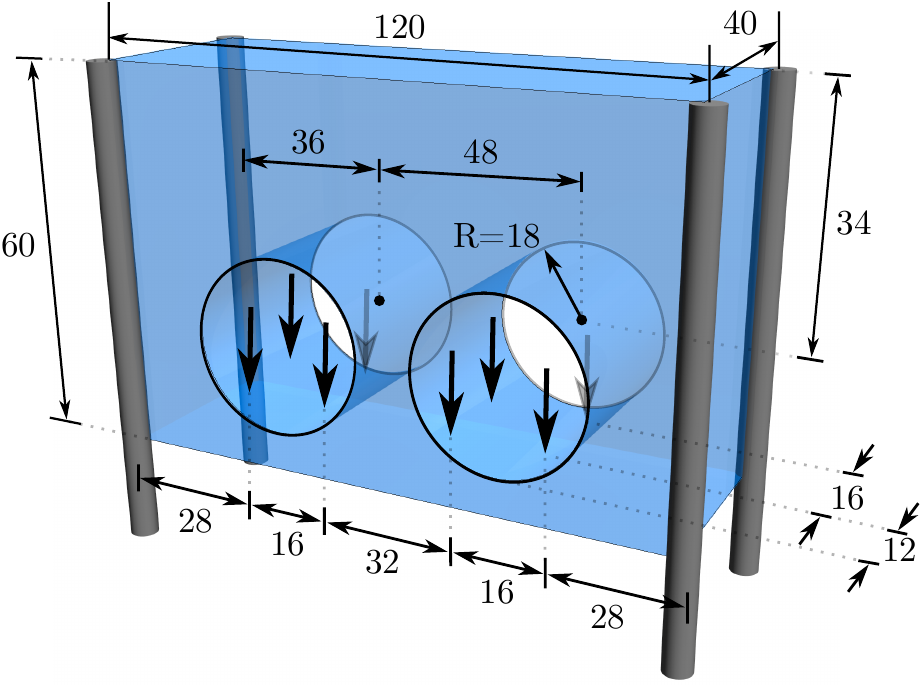}
	\caption{Geometry, boundary conditions and loading of the pipe bracket.}
	\label{fig:pipe_system}
\end{figure}

%
As a structure with a truly three-dimensional load path, the pipe bracket shown in Figure~\ref{fig:pipe_system} is considered.  The design domain has a length, height and width of $120 \times 40 \times 60$ and contains two openings with each of radius of 18 for two pipes passing through the domain. Within the openings each pipe is supported at four points, applying at each support a vertical force of $F=100$. The four vertical outer edges of the design domain are chosen as fixed. The finite element discretisation consists of $120 \times 40 \times 60$ linear hexahedral elements. To represent the two openings in the finite element model the Young's modulus of the elements within the void region is prescribed with~$E_{\text{min}}$.
\begin{figure}[]
	\vspace{0.5em}
	\centering
	\subfloat[][Thresholded geometry with $\eta=0.5$ \label{fig:pipe_top}] {
		\includegraphics[width=0.375\textwidth]{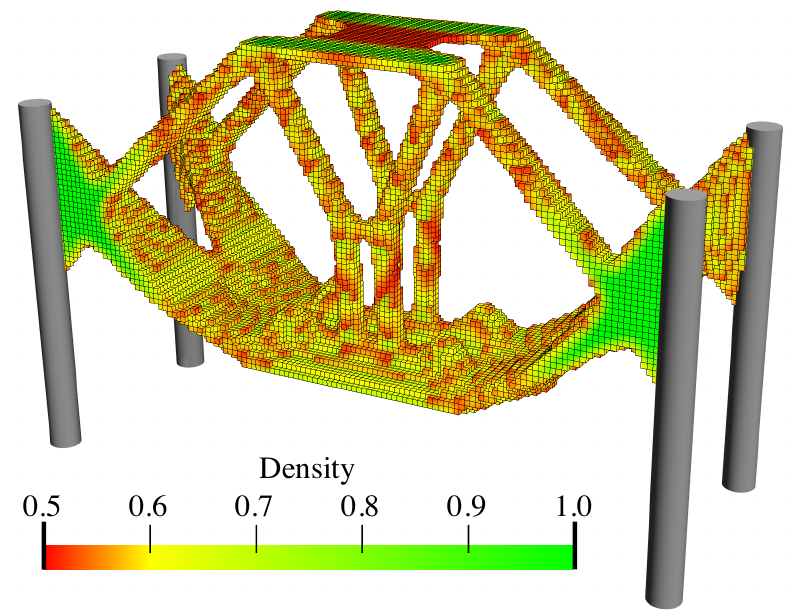}
	}
	\hspace{0.125\textwidth}
	\subfloat[][Voxel chain skeleton \label{fig:pipe_skeleton}] {	
		\includegraphics[width=0.375\textwidth]{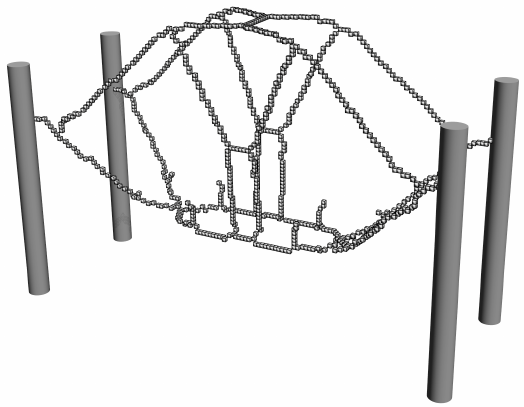}
	}
	\caption{Topology optimised and skeletonised pipe bracket.}
\end{figure}
The maximum volume fraction of the optimised structure is prescribed to be $V_f = 0.1$ and the density filter radius is chosen as $R = 3$. Figure~\ref{fig:pipe_top} depicts the optimised bracket with only the voxels above a relative density $\eta=0.5$. The compliance of the optimised voxel model is \mbox{$J(\hat{\vec \rho})=2.47602$}. As easily recognisable from Figure~\ref{fig:pipe_top}, the optimised structure is a combination of an arch-like and a cable-like structure with vertical tension members between the two openings. 

\begin{figure}[]
	\centering 
	\subfloat[][Size optimised (initial) \label{fig:pipe_frame}] {
		\includegraphics[width=.375\textwidth]{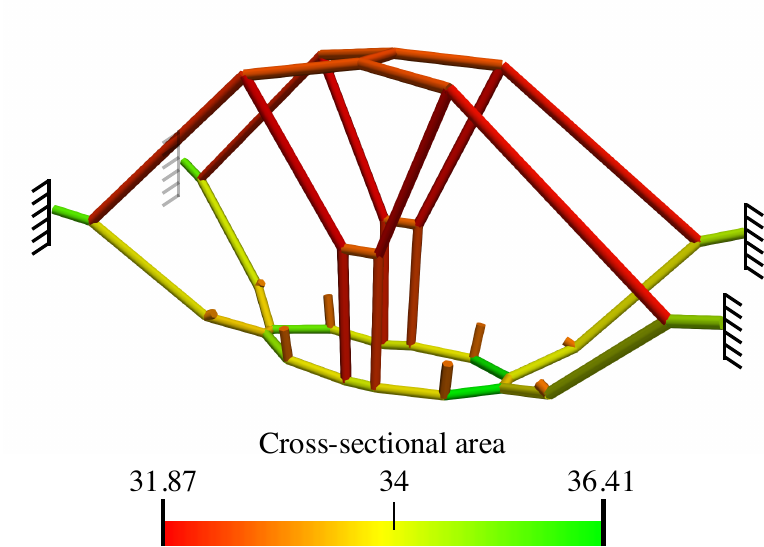}
	}
	\hspace{0.125\textwidth}
	\subfloat[][Size and layout optimised (final) \label{fig:pipe_result}] {
		\includegraphics[width=.375\textwidth]{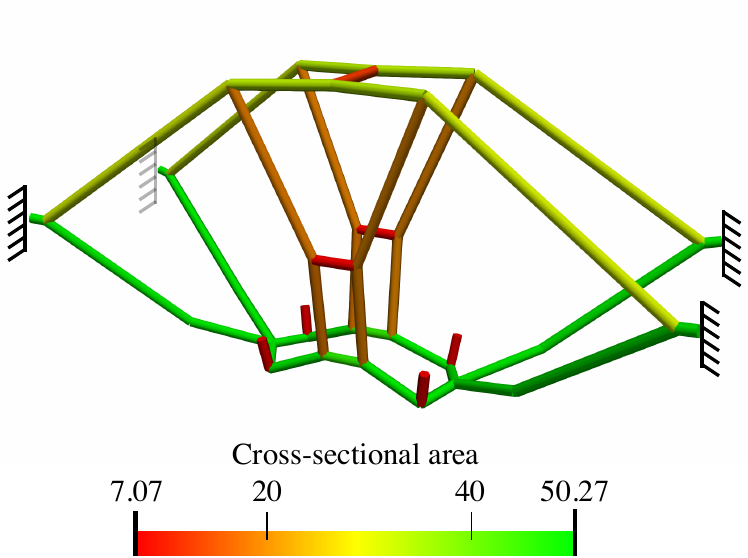}
	}
	\caption{Size and layout optimised pipe bracket.}
\end{figure}

The skeletonisation algorithm yields after $6\times6$ voxel removal steps the voxel chain skeleton in  Figure~\ref{fig:pipe_skeleton}. 
 The skeleton is converted into the structure depicted in Figure~\ref{fig:pipe_frame} by first identifying the 42 joints and then connecting them with 50 straight members. 
 Alternatively, the 50 members can be represented with curved B\'ezier segments, which would lead to a structural frame with curved beam members. This has not been pursued here because such members may lead to an increase in manufacturing costs and are usually not preferred. Initially,  all frame members are assumed to have the same diameter $d=6.296$, giving a total volume of $V = 0.1 \overline V$. 

\begin{figure}[]
	\vspace{0.5em}
	\centering
	\subfloat[][Topology optimisation]
	{
		\includegraphics[scale=0.5]{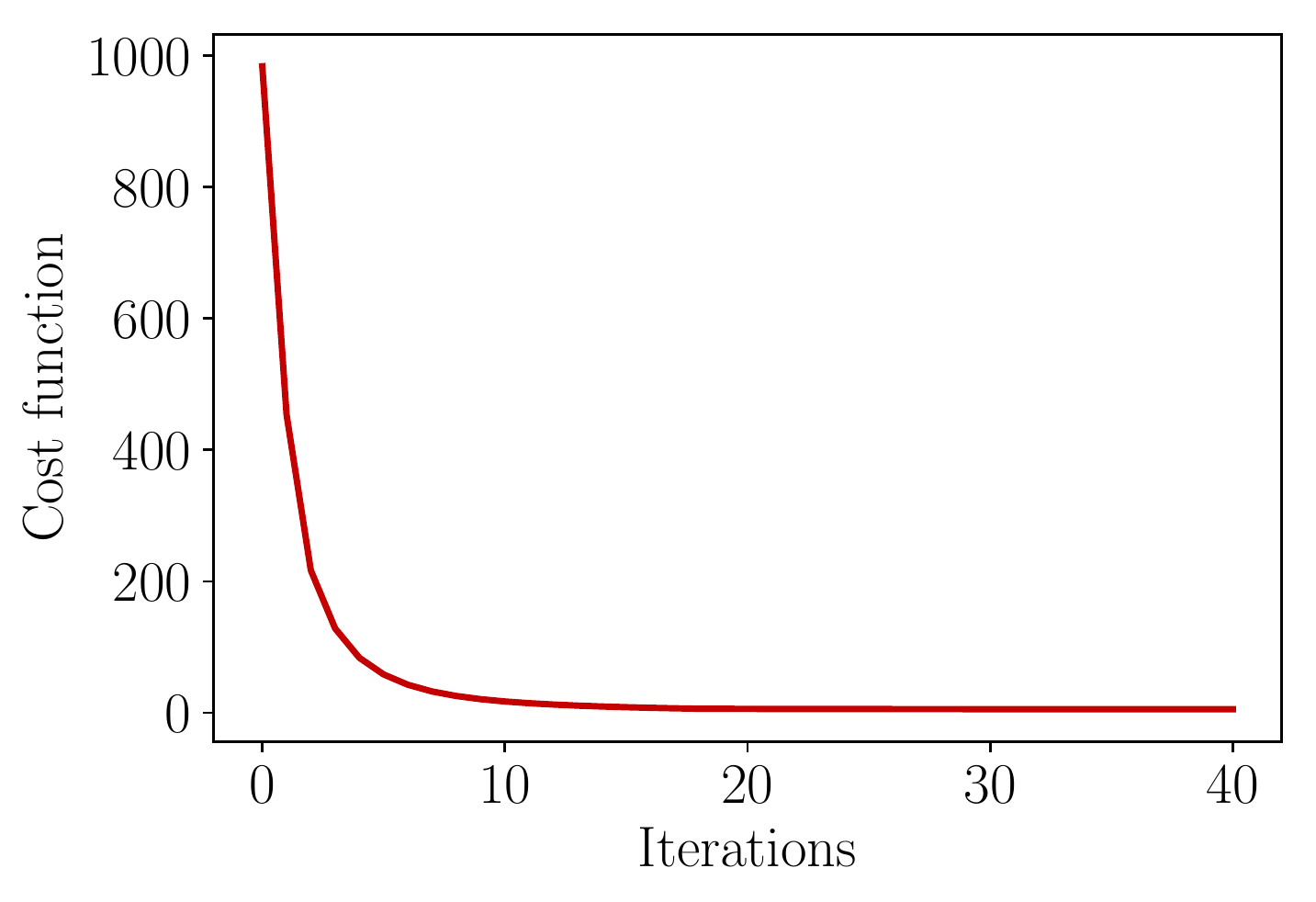}
		\label{fig:pipe_conv_top}
	}
	\subfloat[][Size (S) and layout (L) optimisation]
	{
		\includegraphics[scale=0.5]{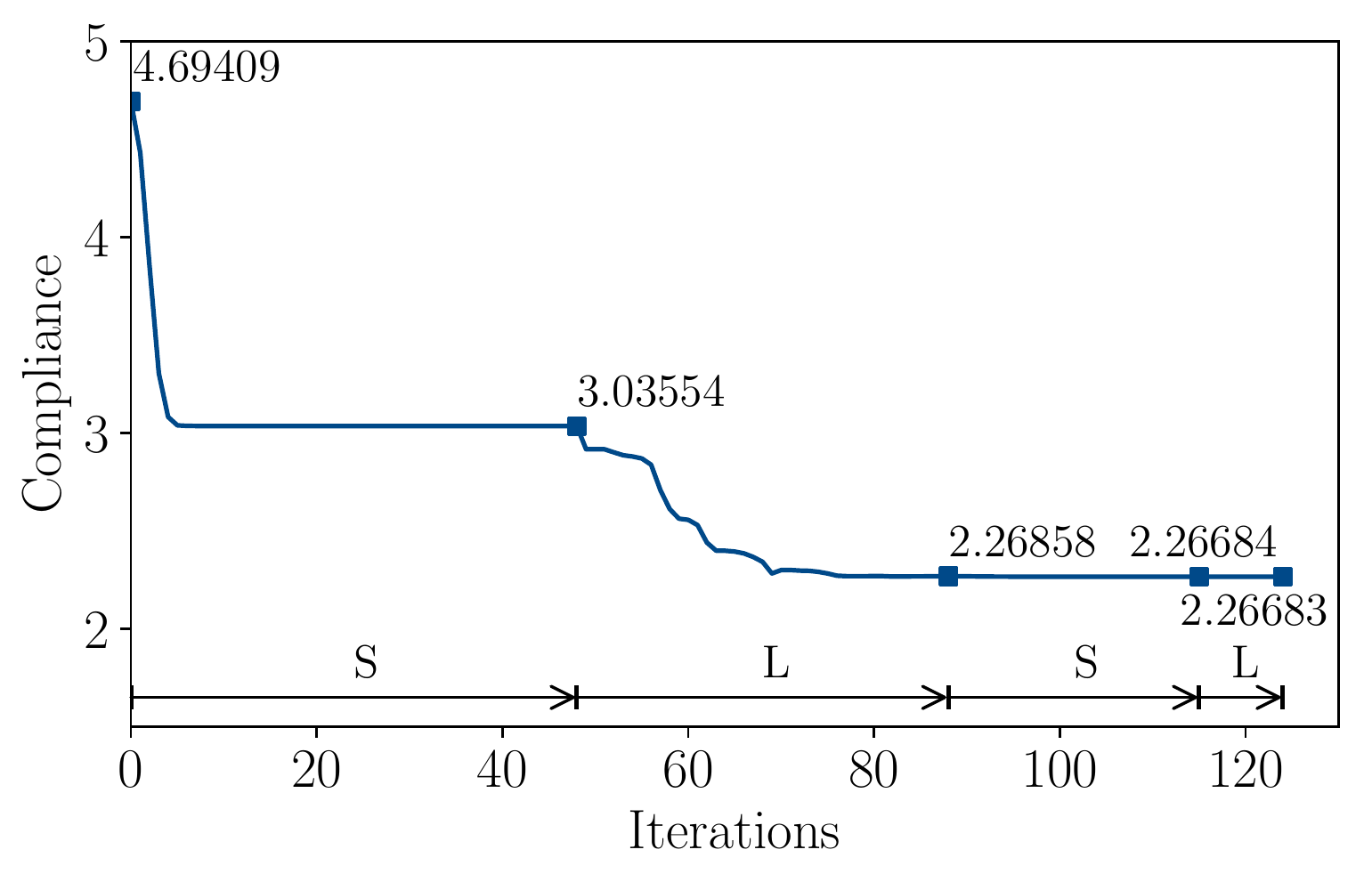}
		\label{fig:pipe_conv_frame}
	}
	\caption{Convergence of the compliance during topology and sequential size and layout optimisation of the pipe bracket.	 Compliance of the topology optimised voxel model is $J(\hat{\vec \rho})=2.47602$. \label{fig:pipe_conv}}
\end{figure}

\begin{figure}[]
	\vspace{0.5em}
	\centering
	\subfloat[][IGES model \label{fig:pipeHolder_iges}]
	{
		\includegraphics[width=0.4\textwidth]{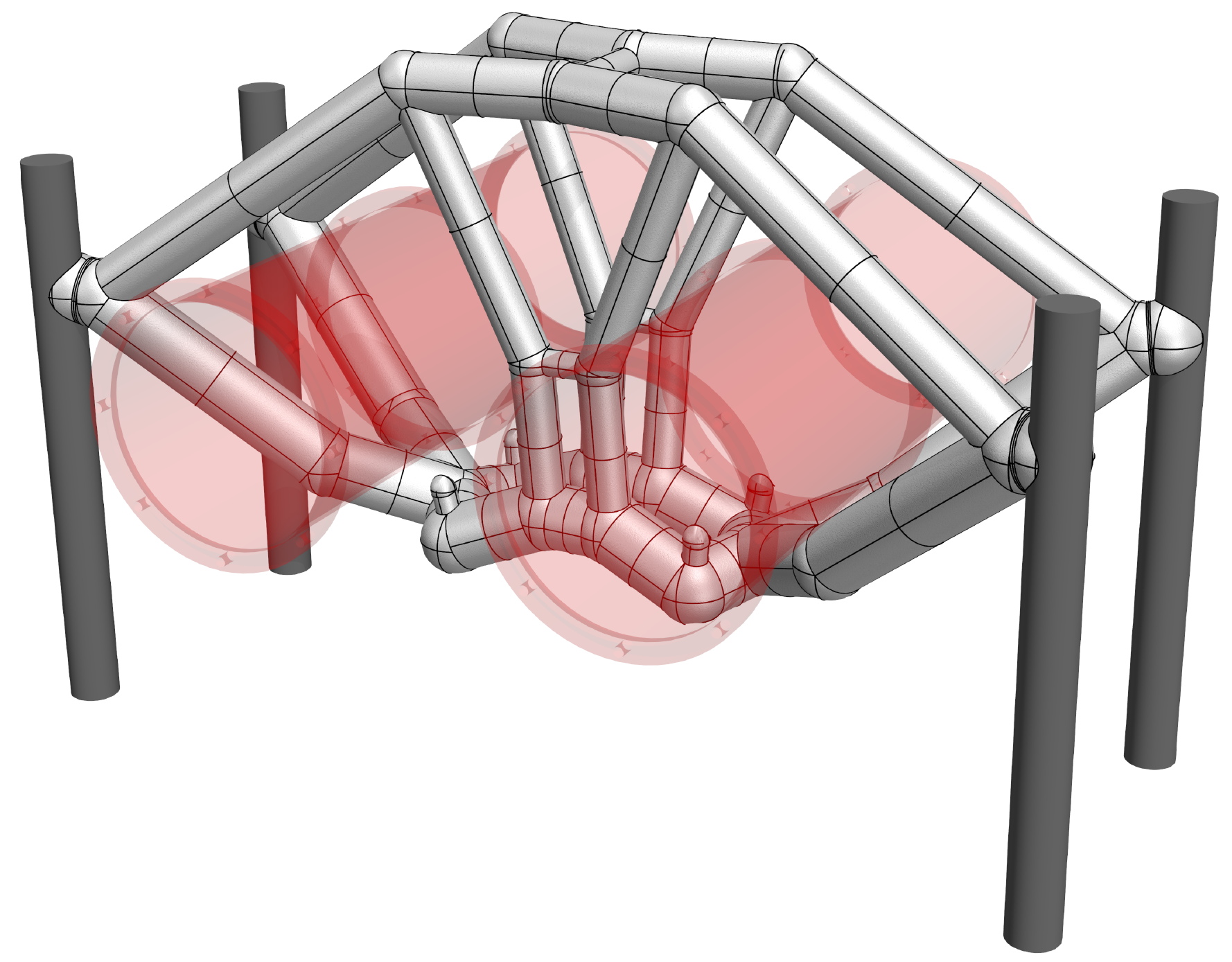}
	}
	\hspace{0.1\textwidth}
	\subfloat[][STL model \label{fig:pipeHolder_stl}]
	{
		\includegraphics[width=0.4\textwidth]{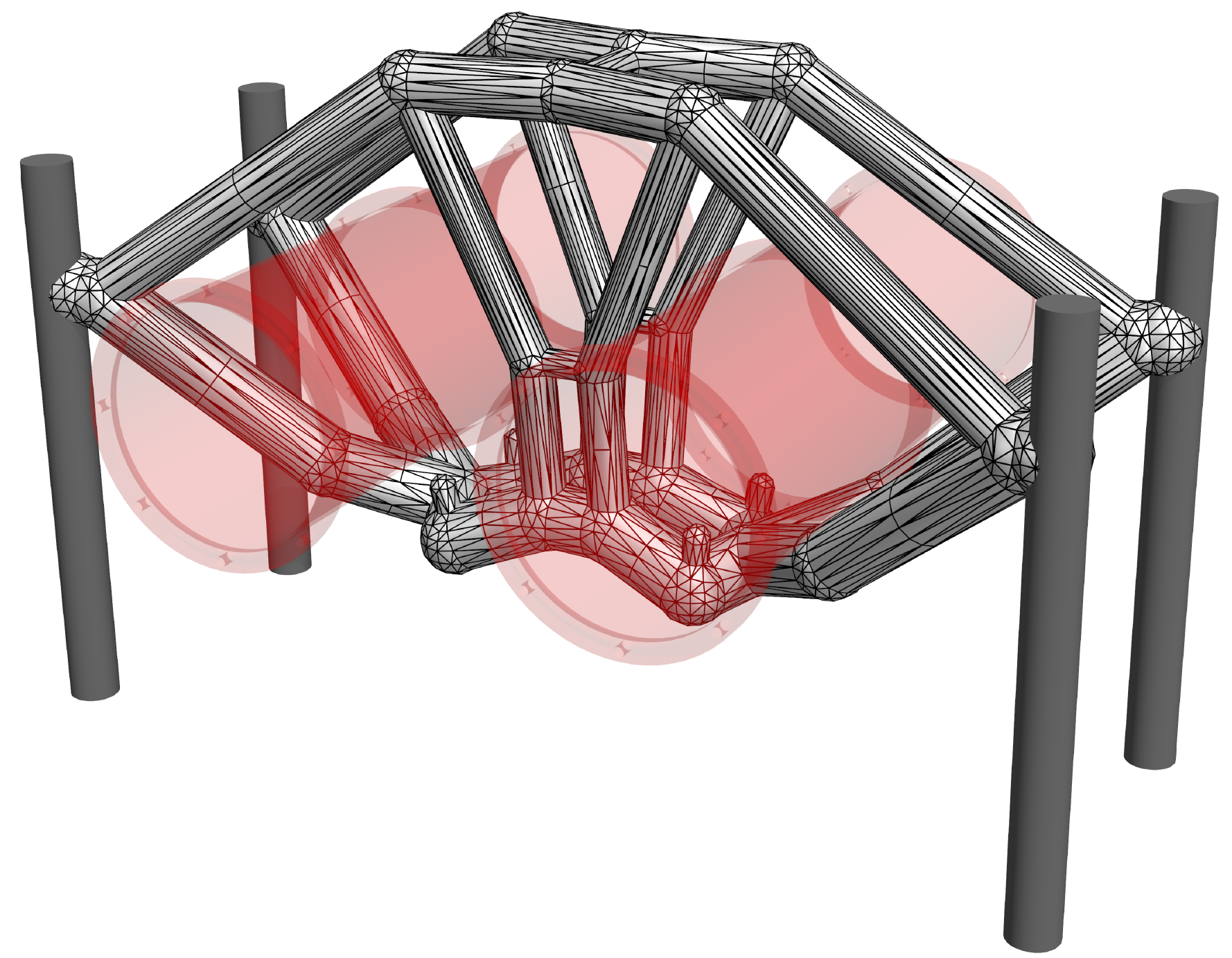}
	}
	\caption{Parametric CAD model of the pipe bracket. \label{fig:pipeHolder_cad}}
\end{figure}

According to Figure~\ref{fig:pipe_conv_frame}, the frame with beams of uniform diameter has a compliance \mbox{$J(\vec d, \, \vec x) = 4.69409$}, which is significantly higher than \mbox{$J(\hat{\vec \rho})=2.47602$} of the voxel model. After the first size optimisation step the compliance is reduced to \mbox{$J(\vec d, \,\vec x) = 3.03554$} and the subsequent layout optimisation step to \mbox{$J(\vec d, \, \vec x)=2.26858$}. In layout optimisation the beams are constrained not to move inside the volume occupied by the two cylindrical pipes. The intersection between a beam and a cylinder is determined by integrating the (positive) signed distance of the cylinder along the beam. It is straightforward to differentiate this constraint integral with respect to nodal coordinates of the beam. The intersection constraints for all the beams are aggregated with the Kreisselmeier-Steinhauser (K-S) function~\cite{martins2005structural}.
Several more steps of size and layout optimisation do not lead to a significant reduction in the cost function. Notice that the obtained final compliance $J(\vec d, \, \vec x)= 2.26683$ is lower than the compliance $J(\hat {\vec \rho})=2.47602$ of the topology optimised voxel model. The final optimised frame structure depicted in Figure~\ref{fig:pipe_result} consists of 36 joints and 44 members. In Figure~\ref{fig:pipeHolder_cad} the solid CAD model of the pipe bracket and a faceted triangular STL mesh exported from FreeCAD are shown.

\clearpage
%
\subsection{Frame-supported plate \label{sec:framePlate}}
\begin{figure}[]
	\centering
	\includegraphics[width=0.475\textwidth]{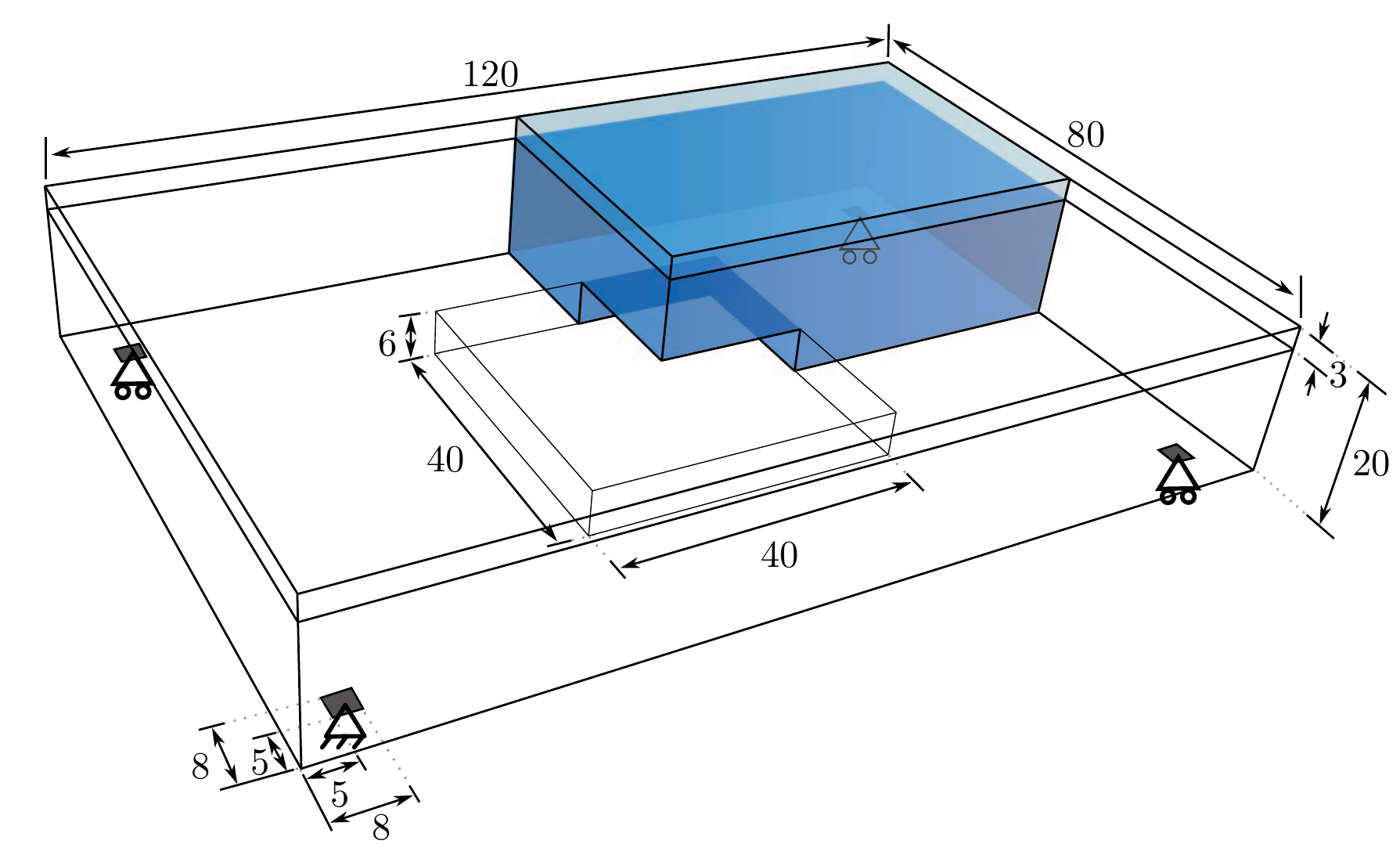}
	\caption{Geometry, boundary conditions and loading of the frame-supported plate. In topology optimisation only one quarter and in size and layout optimisation the entire structure are considered. \label{fig:deck_system}}
\end{figure}

%
In structural design it is very common to combine  a plate- or shell-like skin structure with a frame-like support structure. In such composite structures the skin can be modelled as a standard plate or shell structure, and topology optimisation is applied only in the remaining part of the design domain. The design domain shown in Figure~\ref{fig:deck_system} combines a thin plate-like  skin with a solid bottom part to be optimised. The entire domain is of size $120 \times  80 \times 20$ and contains a small opening of size $ 40 \times 40 \times 6$. The  plate is of thickness $3$  and is subjected to a uniform pressure load of~$1$.  The  four corners of the design domain are vertically supported. Due to symmetry only a quarter of the design domain is considered in topology optimisation. The finite element discretisation consists of  $60 \times  40 \times 23$ linear hexahedral finite elements. Across the height of the design domain there are $6$ elements with a size of $1\times1\times0.5$ in the plate-like top part and $17$ unit sized elements in the bottom part. To represent the small opening in the finite element model the Young's modulus of the relevant elements is prescribed as~$E_\text{min}$.  
\begin{figure}[t]
	\vspace{0.5em}
	\centering 
	\subfloat[][Thresholded geometry with $\eta=0.55$ \label{fig:deck_top} ]{
		\includegraphics[width=0.375\textwidth]{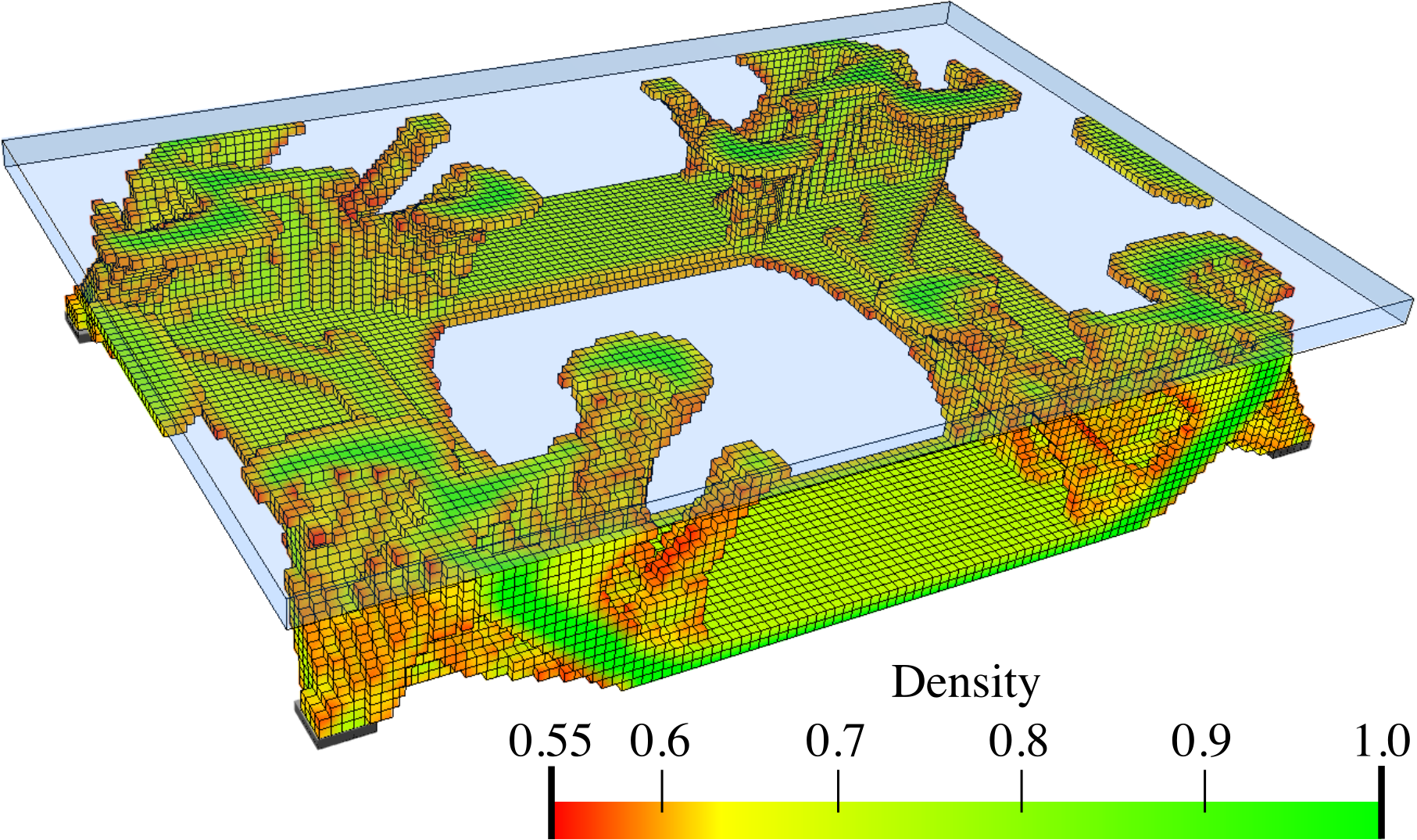}
	}
	\hspace{0.125\textwidth}
	\subfloat[][Voxel chain skeleton  \label{fig:deck_skeleton}]{
		\includegraphics[width=0.375\textwidth]{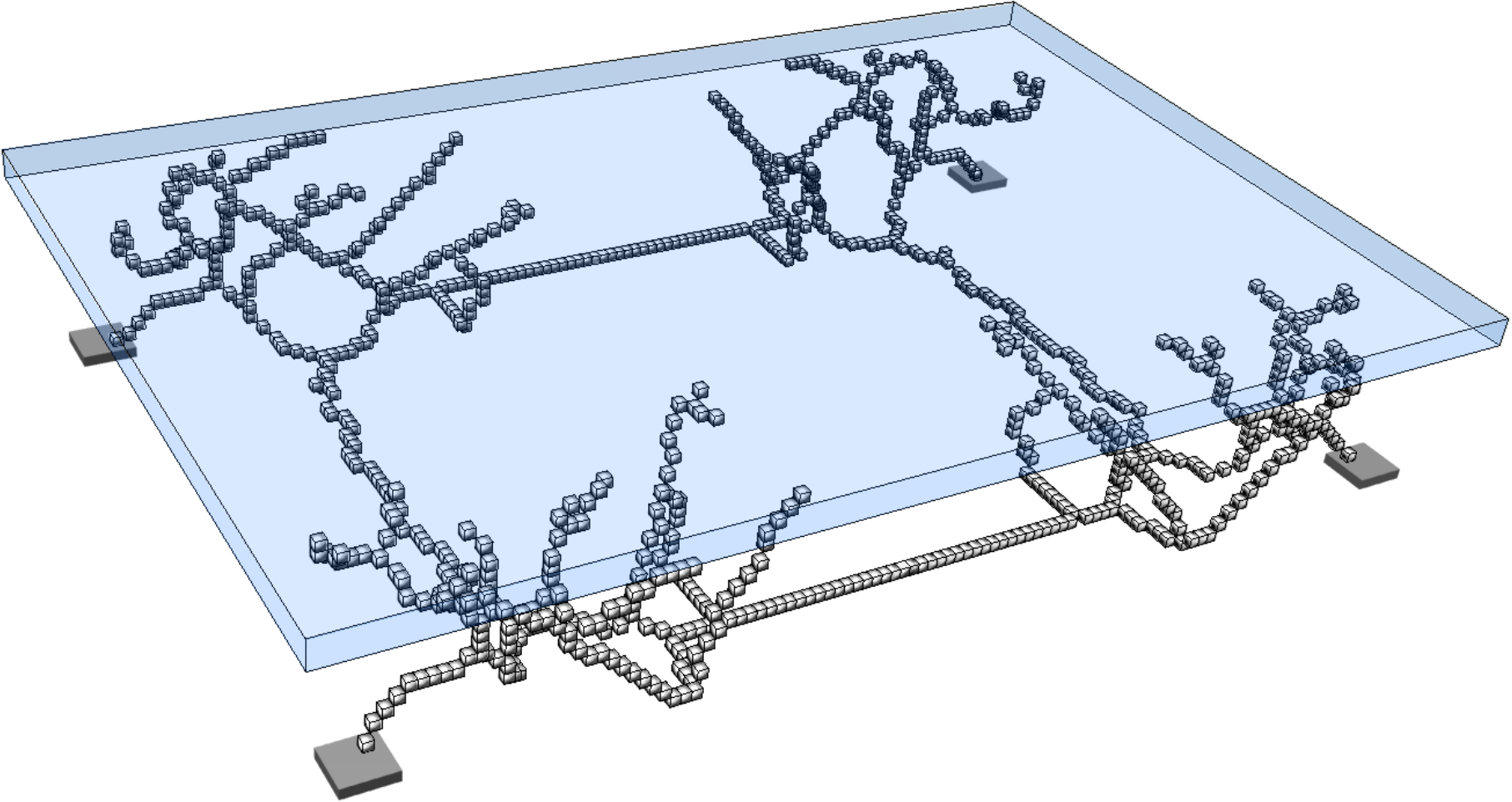}
	}
	\caption{Topology optimised and skeletonised frame-supported plate,}
\end{figure}

\begin{figure}[h!]
	\centering 
	\subfloat[][Size optimised (initial) \label{fig:deck_frame}]{
		\includegraphics[width=0.375\textwidth]{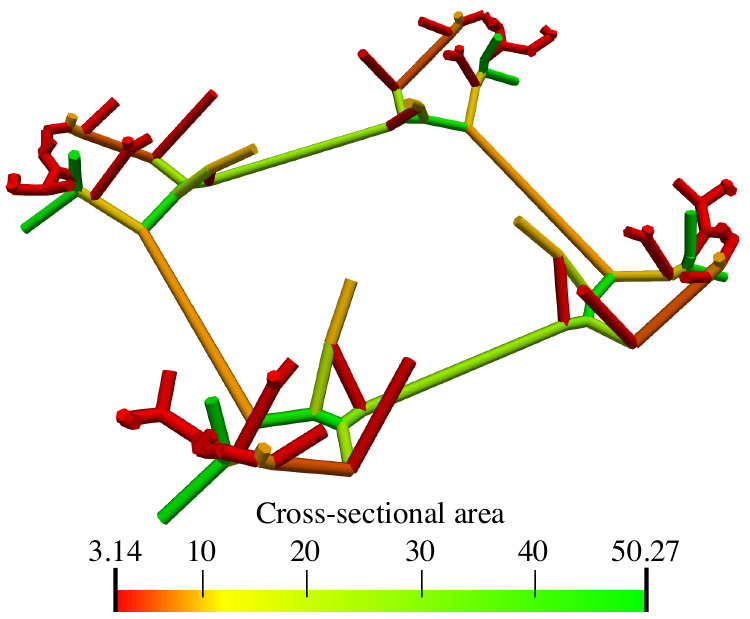}
	}
	\hspace{0.125\textwidth}
	\subfloat[][Size and layout optimised (final) \label{fig:deck_result}]{
		\includegraphics[width=0.375\textwidth]{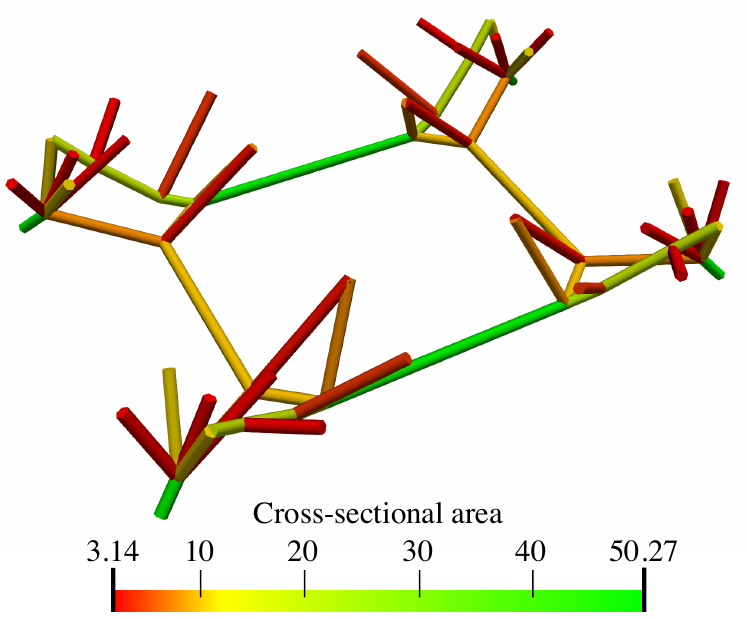}
	}
	\caption{Size and layout optimised frame-supported plate. The plate on the top is not shown for clarity. \label{fig:sizeLayoutOptimised}}
	\vspace{0.5em}
	\centering
	\subfloat[][Topology optimisation \label{fig:deck_conv_top}]
	{
		\includegraphics[scale=0.5]{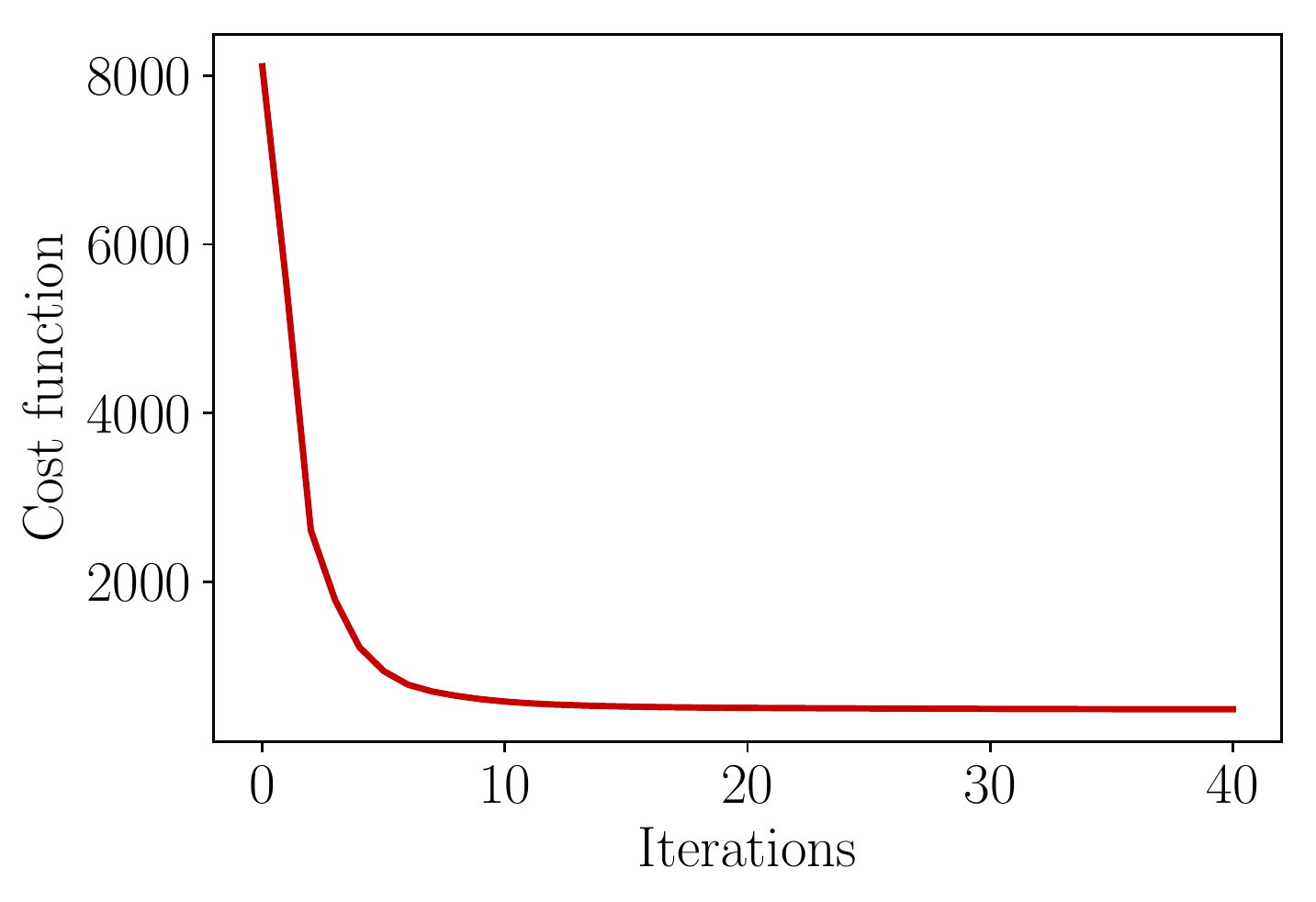}
	}
	\subfloat[][Size (S) and layout (L) optimisation \label{fig:deck_conv_frame}]
	{
		\includegraphics[scale=0.5]{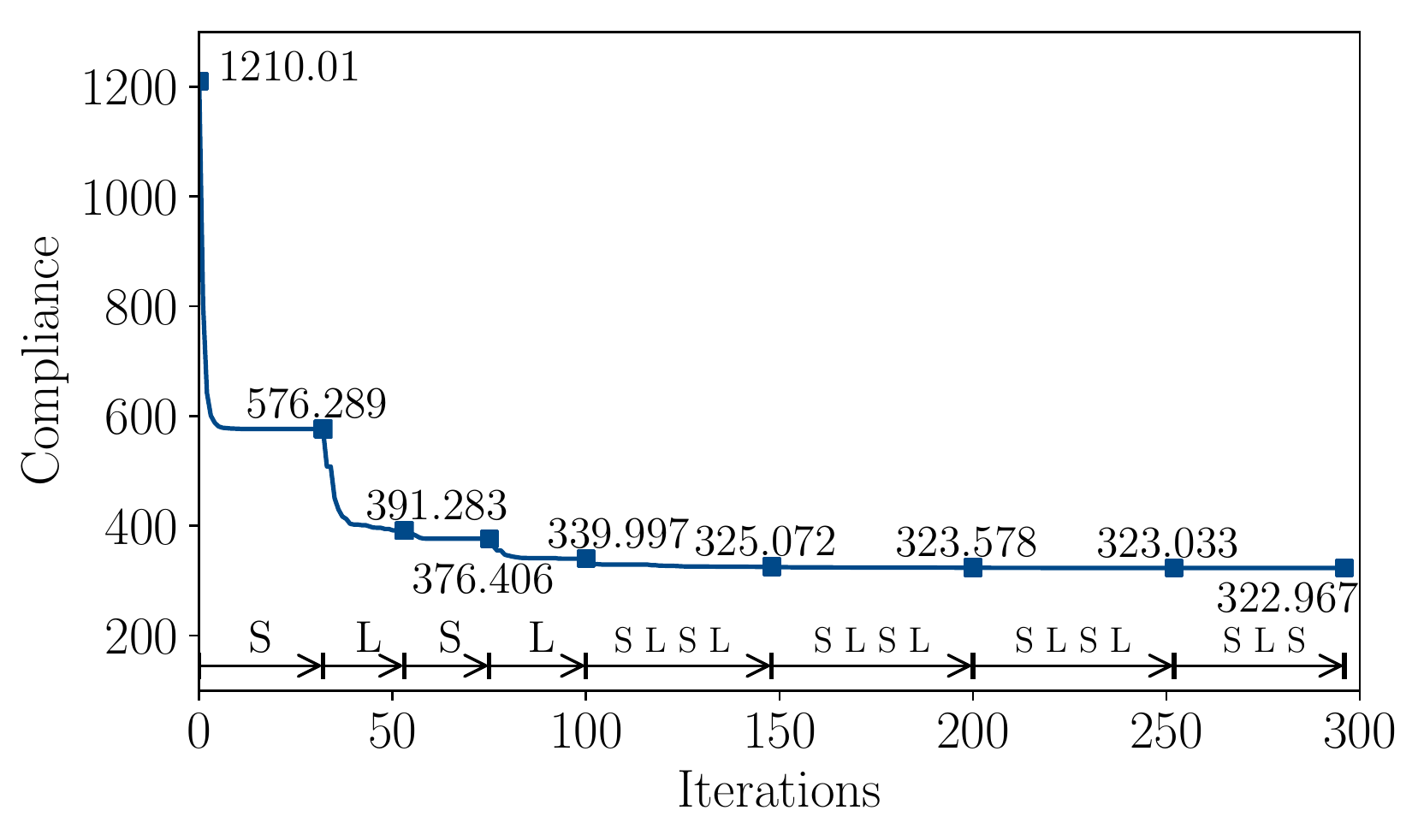}
	}
	\caption{Convergence of the compliance during topology and sequential size and layout optimisation of the frame-supported plate. Compliance of the topology optimised voxel model is $J(\hat{\vec \rho})=364.596$. \label{fig:deck_conv}}
%
	\vspace{0.5em}
	\centering
	\subfloat[][STL model \label{fig:deck_stl}]
	{
		\includegraphics[width=0.4\textwidth]{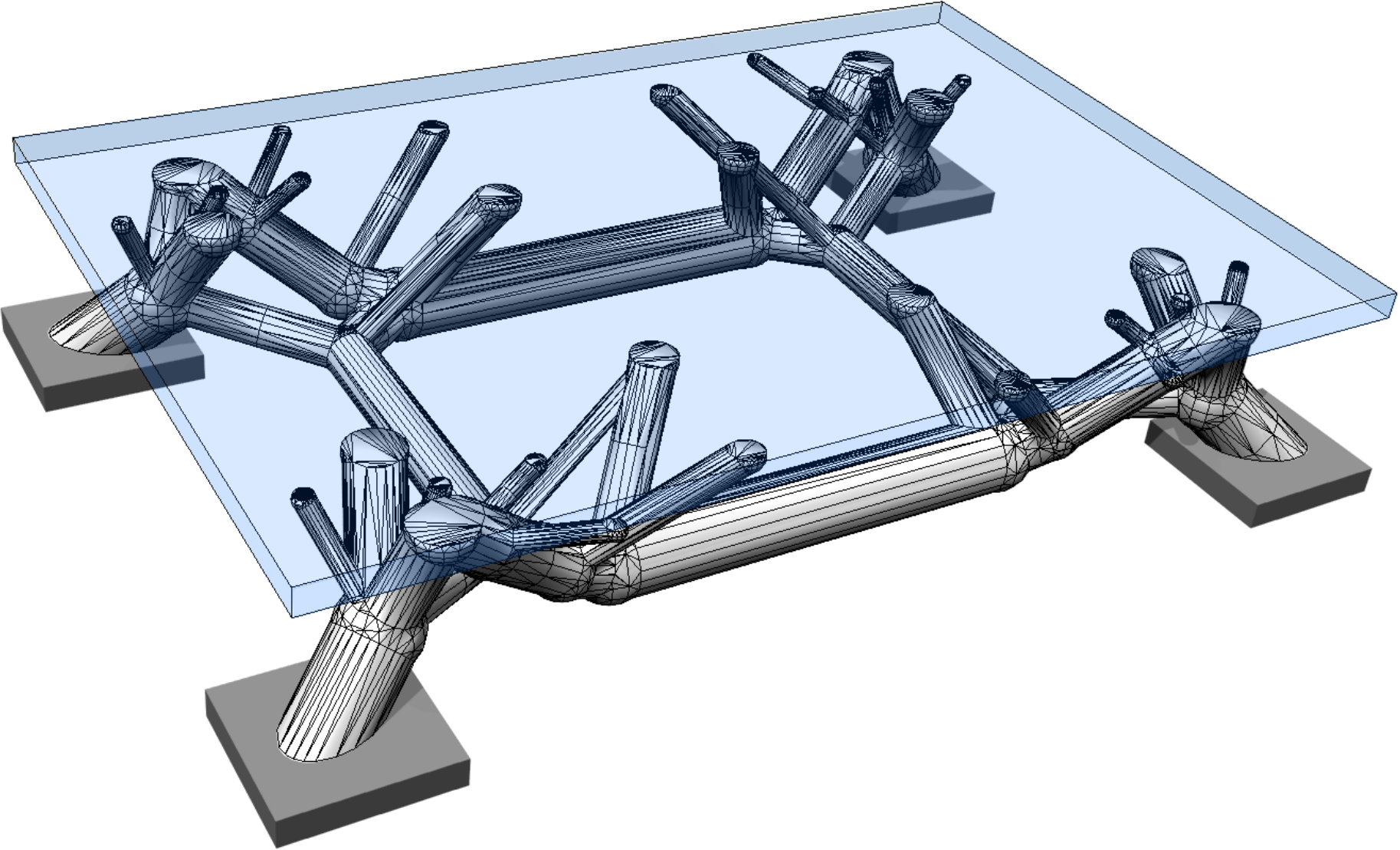}
	}
	\hspace{0.1\textwidth}
	\subfloat[][IGES model \label{fig:deck_iges}]
	{
		\includegraphics[width=0.4\textwidth]{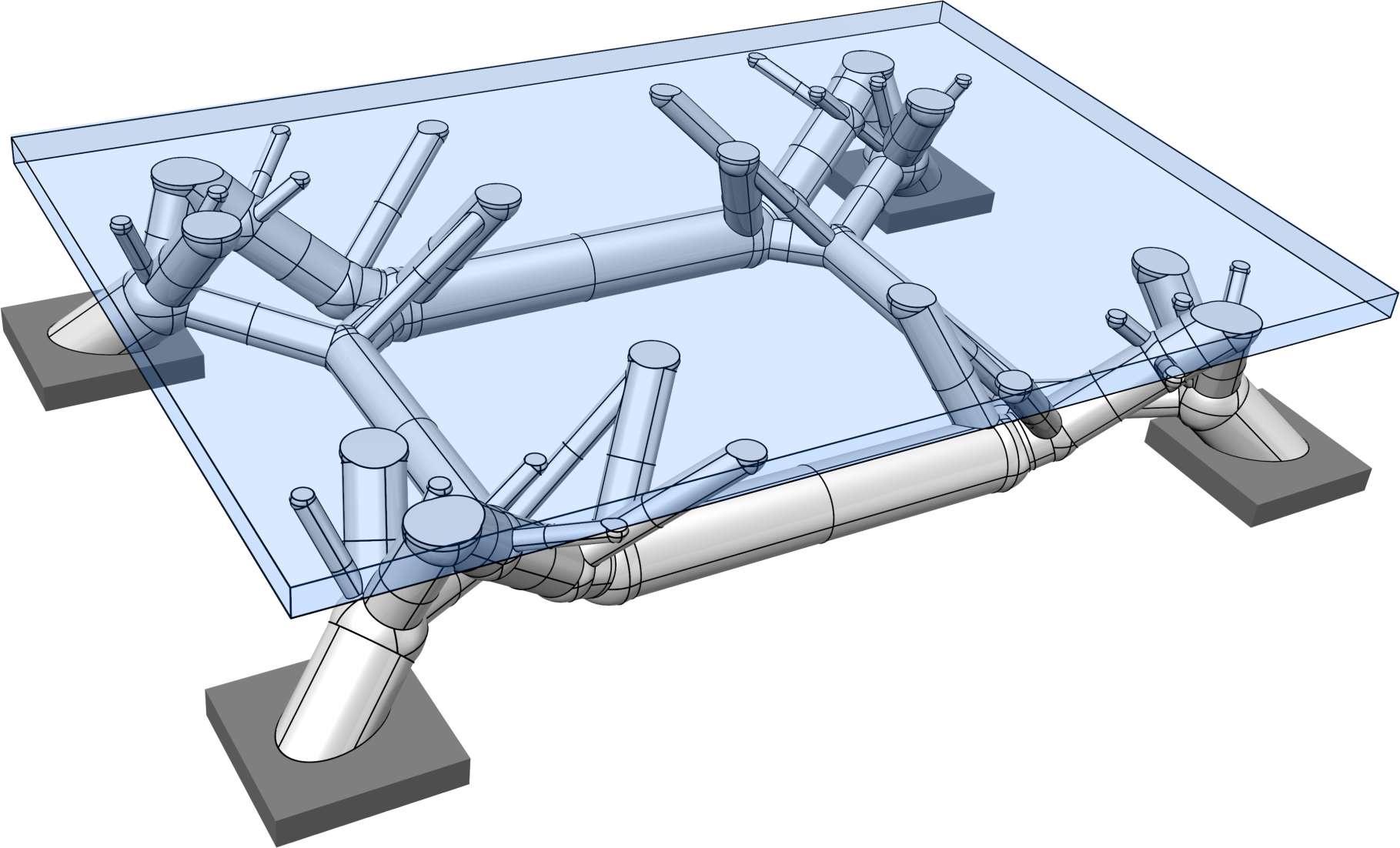}
	}
	\caption{Parametric CAD model of the frame-supported plate. \label{fig:deck_cad}}
\end{figure}

In topology optimisation, the maximum volume fraction is prescribed to be $V_f = 0.25$ and the density filter radius is chosen as $R = 1.5$. The optimised structure with only the voxels above a relative density $\eta = 0.55$ is shown in Figure~\ref{fig:deck_top} and its voxel chain skeleton in Figure~\ref{fig:deck_skeleton}. Although the voxels in the top part of the design domain are present during optimisation and skeletonisation, they are tagged as non-removable. The voxel chain skeleton is converted to the frame structure shown in Figure~\ref{fig:deck_frame}. The structural model also contains a top plate not shown in Figure~\ref{fig:sizeLayoutOptimised}.  The plate is modelled as a Kirchhoff--Love plate and is discretised with  $4\times4$ quadrilateral elements and Catmull-Clark subdivision basis functions. 
The frame  has 124 joints and 136 members. Initially, all the members are assumed to have the same diameter of $d=5.010$ which gives a total frame volume of $V = 4800$. The coupling between the plate and the frame is achieved with Lagrange multipliers as described in~\cite{xiao2019interrogation}. The joints between the plate and frame can transfer forces but not moments.

According to Figure~\ref{fig:deck_conv_frame}, the compliance of the frame-supported plate with members of uniform diameter is \mbox{$J(\vec d, \, \vec x) =1210.01$}, which reduces after  several sequential size and layout optimisation steps to~\mbox{$J(\vec d, \, \vec x) =322.967$}. During the layout optimisation, the positions of the joints between the frame and plate are fixed. For comparison,  the compliance of the topology optimised voxel model was $J(\hat{\vec \rho})=364.596$.  
The final optimised frame structure contains 56 joints and 64 members, see Figure~\ref{fig:deck_result}. Its solid CAD model and faceted triangular STL mesh exported from FreeCAD are depicted in Figure~\ref{fig:deck_cad}.

%
\section{Conclusions \label{sec:conclusions}}
%
The automated conversion of topology optimised geometries into compact parametric CAD models is a missing link in the wide-scale industrial adoption of optimisation. We introduced a fully automated workflow for synthesising a structurally-sound parametric CAD model from a voxelised solid-void binary image obtained with topology optimisation. Currently, there is no commercial design system, such as Altair Inspire, Abaqus CAE or similar, which offers such a fully automated CAD model generation capability. Homotopic skeletonisation of three-dimensional images is a key component of our proposed workflow, which is an extensively well-studied topic in digital topology. The resulting voxel chain skeleton is interpreted as a weighted undirected graph and processed with standard graph algorithms to yield a frame model. Both the skeletonisation and graph algorithms operate on topology rather than geometry and are not subject to floating-point errors, which makes them exceedingly robust. The so-obtained structural frame models are not optimal because the skeletonisation and graph algorithms do not take into account any mechanical considerations. However, as shown in the numerical examples, after several alternating steps of size and layout optimisation, the structural frames can achieve a cost function value similar to the original topology optimised geometry. The frame model is converted into a compact CAD solid model by recursively combining primitive shapes using boolean operations. Although we used in this paper only cylinders and spheres as shape primitives, it is straightforward to use other primitives. There is also scope in improving the CAD solid model itself, expressly, by using the recently proposed quadratic of revolution (quador) representation for lattices and frames~\cite{gupta2018quador,gupta2019exact,cirak2020adding}. A quador CAD model consists only of conic curves and surfaces, which greatly simplifies and accelerates frequently required geometry queries, like closest point computations or slicing,  critical to meshing and manufacturing process planning. 

The generated parametric CAD solid model and its compact CSG tree representation can be edited to refine the optimised design or to combine it with other components into a product.  It is worth reminding that in industrial practice, as supported by user studies~\cite{bradner2014parameters}, geometries obtained from optimisation are often the starting point for design exploration and not the endpoint. This usually requires the ability to edit the optimised geometries in a CAD system. A high-level compact representation is also useful for taking into account geometric manufacturing constraints, like minimum thickness or slope or length of the members~\cite{liu2016survey}. Beyond geometry editing, the frame model may also have advantages in structural analysis of the CAD model.  The frame model makes it, for instance, easier to check the structure for buckling and inelastic deformations. As a low-order model, it can be analysed orders of magnitude faster than a three-dimensional solid model. This opens up the possibility of instant analysis of CAD models as it has become recently available in, e.g., Ansys Discovery Live or Creo Simulation Live. 
\begin{figure}
	\centering 
	\subfloat[][Voxel chain skeleton \label{fig:pipeHolderSurfA}] {	
		\includegraphics[width=0.4\textwidth]{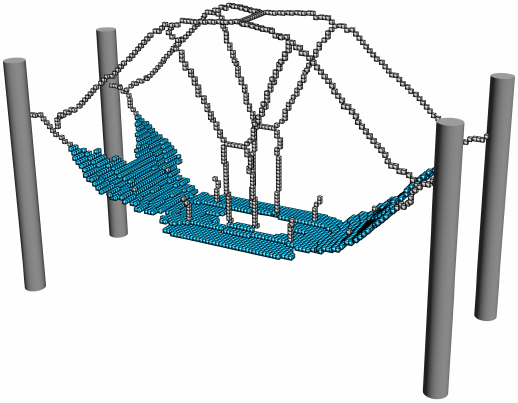}
		}
	\hspace{0.1\textwidth}	
	\subfloat[][IGES model\label{pipeHolderSurfB}] {
		\includegraphics[width=0.4\textwidth]{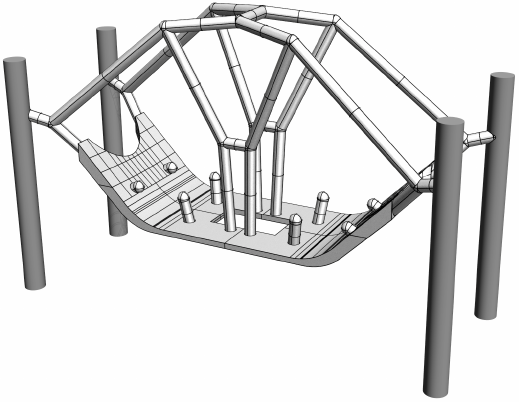}
	}
	\caption{Surface-preserving skeletonisation and CAD model of the topology optimised pipe bracket in Figure~\ref{fig:pipe_top}.  
	 \label{fig:pipeHolderSurf}}
\end{figure}

In closing, one particular extension of the presented approach is worth mentioning.  We assumed that the topology optimised voxel model can be approximated with tubular members. However, in three-dimensional optimisation problems, large flat or curved surfaces may also appear during optimisation. The introduced skeletonisation algorithm in Section~\ref{sec:digTopSkeleton} will reduce even those into a network of one-voxel-wide chains. As suggested in Lee et al.~\cite{lee1994building}, the algorithm can be modified to preserve one-voxel-wide surfaces which appear during skeletonisation. This is achieved by not deleting the so-called surface points/voxels, which are defined based on the arrangement of solid voxels in an octant. In Figure~\ref{fig:pipeHolderSurf} the topology optimised pipe bracket has been skeletonised with the modified skeletonisation algorithm. The curved surface at the bottom is preserved. In the structural model, this surface can be modelled as a thin-shell and in the CAD model it can be easily converted into a solid by providing a thickness. Currently, we do not have an automated process to extract a shell surface from the skeletonised voxel model, but this seems to be possible and is a promising direction for future research.

\appendix 

%
\section{Member stiffness matrix} \label{sec:appendix}
%

\begin{figure}[b]
    \centering
	\subfloat[Displacement degrees of freedom \label{fig:translationDisp}]
	{
		\includegraphics{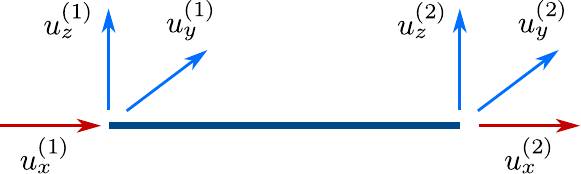}
	}
	\hspace{0.1\textwidth}
	\subfloat[Rotation degrees of freedom \label{fig:rotationDisp}]
	{
		\includegraphics{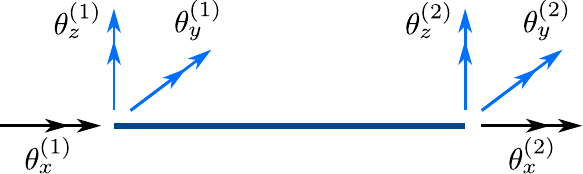}
	}
    \caption{Degrees of freedom of a Timoshenko beam element (coupled degrees of freedom are depicted in the same colours).}
    \label{fig:beamDofs}
\end{figure}

We model the members of the structural frame as Timoshenko beams. As mentioned, the members are assumed to be straight and have uniform cross-sections along their lengths. Hence, each member can be approximated with a single beam finite element and its stiffness matrix can be derived with standard structural analysis techniques from undergraduate textbooks, see e.g.~\cite{megson2016aircraft}. The stiffness matrix we use takes into account axial stretch, transversal shear, bending and torsion effects. The local stiffness matrix~$\vec K_i^l$ of a beam $i$ is derived in a local coordinate system in which the beam axis is aligned with the~$x_i^l$ axis. Each of the two end nodes of the beam have three displacement and three rotation degrees of freedom, see Figure~\ref{fig:beamDofs}.  To ease the notation, we write in the following for the local coordinate axes only $x$, $y$ and $z$. 

The element stiffness matrix of the beam is composed out of independent axial, bending and torsion stiffness matrices. The axial stiffness matrix corresponding to the displacement degrees of freedom~$u_x^{(1)}$ and~$u_x^{(2)}$ is given by 
\begin{equation} \label{eq:stretchStiffness}
	 \frac{EA}{L}
    \begin{pmatrix*}[r]
    1 & -1 \\
    -1 & 1
    \end{pmatrix*} \, ,
\end{equation}
where $E$, $A$ and $L$ denote the Young's modulus, cross-sectional area and length. 

The bending stiffness matrix corresponding to the displacement degrees of freedom $u_{y}^{(1)}$,  $u_{z}^{(1)} $,  $u_{y}^{(2)}$  and $u_{z}^{(2)}$ and the rotational degrees of freedom  $\theta_{y}^{(1)}$, $ \theta_{z}^{(1)}$, $ \theta_{y}^{(2)}$ and $ \theta_{z}^{(2)}$ is given by
\begin{align} \label{eq:bendingStiffness}
\renewcommand\arraystretch{2}
    \begin{pmatrix*}[c]
    \frac{12EI_z}{(1 + b_z)L^3} & 0 & 0 & \frac{6EI_z}{(1 + b_z)L^2}  & -\frac{12EI_z}{(1 + b_z)L^3} & 0 & 0 & \frac{6EI_z}{(1 + b_z)L^2} \\
      & \frac{12EI_y}{(1 + b_y)L^3} & -\frac{6EI_y}{(1 + b_y)L^2}  & 0 & 0 & -\frac{12EI_y}{(1 + b_y)L^3} & -\frac{6EI_y}{(1 + b_y)L^2} & 0 \\
      &  & \frac{(4 + b_y)EI_y}{(1 + b_y)L} & 0 & 0 & \frac{6EI_y}{(1 + b_y)L^2} & \frac{(2 - b_y)EI_y}{(1 + b_y)L} & 0 \\
      &   &   & \frac{(4 + b_z)EI_z}{(1 + b_z)L} & -\frac{6EI_z}{(1 + b_z)L^2} & 0 & 0 & \frac{(2 - b_z)EI_z}{(1 + b_z)L} \\
      &   &   &  & \frac{12EI_z}{(1 + b_z)L^3} & 0 & 0 & -\frac{6EI_z}{(1 + b_z)L^2} \\
      &   & \textit{sym.}  &  &   & \frac{12EI_y}{(1 + b_y)L^3} & \frac{6EI_y}{(1 + b_y)L^2} & 0 \\
      &   &   &  &   &  & \frac{(4 + b_y)EI_y}{(1 + b_y)L} & 0 \\
      &   &   &  &   &  &   & \frac{(4 + b_z)EI_z}{(1 + b_z)L}
    \end{pmatrix*}  \, ,
\end{align}
where $I_y$ and $I_z$ are the second moments of area about the~$y$ and~$z$ axis. The two dimensionless factors $b_y$ and $b_z$ take into account the transversal shear and  are defined according to $b = 12EI/(\kappa GAL^2)$, with $G$ the shear modulus and $\kappa$ the shear correction factor.  In case of a circular cross section we have $I_y = I_z = I$, $b_y = b_z = b$ and $\kappa = 0.9$. Note that setting~$b_y = b_z = 0$  in~\eqref{eq:bendingStiffness} gives the stiffness matrix of an Euler-Bernoulli beam. 

Finally,  the torsional stiffness matrix corresponding to the rotational degrees of freedom $\theta_x^{(1)}$ and $\theta_x^{(2)}$  is given by
\begin{equation} \label{eq:torsionStiffness}
    \frac{GJ}{L}
    \begin{pmatrix*}[r]
    1 & -1 \\
    -1 & 1
    \end{pmatrix*} \, ,
\end{equation}
where $J$ is the torsion constant of the cross section.  

The local stiffness matrix~$\vec K_i^l$ of an element~$i$ is obtained by suitably assembling the three stiffness matrices~\eqref{eq:stretchStiffness}, 
\eqref{eq:bendingStiffness} and~\eqref{eq:torsionStiffness} into a $12\times 12$ matrix. This matrix in the local coordinate system $x_i^l$, $y_i^l$ and $z_i^l$ is transformed into a stiffness matrix~$\vec K_i$ in the global coordinate system $x$, $y$ and $z$ according to~\eqref{eq:elmStiffness}.  
%
\section{Algorithms} \label{sec:appendixAlg}
%
The proposed approach essentially relies on the Algorithms~\ref{alg:skeletonisation} and~\ref{alg:extractGraph} mentioned in Sections~\ref{sec:skeletonisation} and~\ref{sec:graphModel}, respectively. The input of Algorithm~\ref{alg:skeletonisation} for skeletonisation consists of the binary voxel model with $\set V = \set V_s \cup \set V_ e$ and the tagged non-removable voxels~$\set V_t$. In each iteration (loop starting line~\ref{alg1:innerLoop}) a layer of border voxels in one of the six coordinate directions is removed. Within each iteration first all potentially removable voxels are determined and then removed (loops starting lines \ref{alg1:innerLoop} and~\ref{alg1:innerLoop2} respectively). Only voxels which have an empty neighbour, are not the end point of a chain, are a simple point and are not tagged as non-removable can be potentially removed. According to~\eqref{eq:changeCrit} a voxel~$v$ is a simple point if its removal does not change the Euler characteristic and the number of connected components of its 26-neighbourhood $\set N_{26}(v)$. We determine the change of the Euler characteristic with a look-up table~\cite{lee1994building,homann2007,post2016fast} and the number of connected components using the Boost Graph Library~\cite{siek2002boost}. The inner loop starting line~\ref{alg1:innerLoop2} is necessary because each voxel is present in 27 26-neighbourhoods and the removal of one voxel can potentially render any one of the neighbouring 26 voxels non-removable. 

\DontPrintSemicolon
\SetAlgoNoLine
\LinesNumbered
\begin{algorithm}[h!]
\caption{Skeletonise \label{alg:skeletonisation}}
\KwIn{ $\set V$, $\set V_s$,  $\set V_t $  \tcp*{binary voxel model and tagged voxels}}
%
$c = 0$  \tcp*{auxiliary counter}

%
%
\While {$c \neq 6$}{
	$c  \leftarrow 0$\\
	\For (\tcp*[f]{loop over all coordinate directions} \label{alg1:innerLoop}){ $i = 1$  \KwTo $6$ }  {   
		$\set D = \emptyset $  \tcp*{potentially deletable voxels}
		\For (\label{alg1:innerLoop1}) {$v\in\set V$}{
			\lIf{ $ \left ( \begin{tabular}{@{\hspace*{0em}}l@{}}   \label{alg1:if}
      	 		       $v \in \set V_s$ {\normalfont \textbf{and}} \\
		      	       {\normalfont neighbour($i, v$) $\notin V_s $  \textbf{and}} \\
		 	       {\normalfont isNotEndPoint $(v) $  \textbf{and} }\\  
		 	       {\normalfont isSimplePoint$(v) $  \textbf{and} }\\  
		    	       $v \notin \set V_t$\\   
    			       \end{tabular} \right ) $}{	
			  	 $\set D \leftarrow \set D \cup \{ v \}$
			}		
		}
		$n_s  =  \vert  V_s \vert $ \\
		\For (\tcp*[f]{delete solid voxels} \label{alg1:innerLoop2}) {$v\in\set  D$}{
		\lIf{\normalfont isSimplePoint$(v)$ }{ 
			$\set V_s \leftarrow \set V_s \setminus \{ v\}$  }
		}
		
		\lIf{$n_s = |V_s|$}{ $c \leftarrow c +1$ } 
	}
} 
\KwOut{$\set V_s$  \tcp*{voxel skeleton}}
\end{algorithm}

The obtained voxel skeleton is converted with Algorithm~\ref{alg:extractGraph}  into a weighted undirected graph. Each joint voxel with \mbox{$| \set N_{26}(v) | \neq 2 $} or tagged voxel~$v \in \set{V}_t$ becomes a node of the graph. The edges are obtained by starting from a joint voxel~$v$ and marching along all its neighbours~$\set N_{26}(v)$, in turn, until a second joint or tagged voxel is reached (loop starting line~\ref{alg2:march}). The edge weight is  obtained by incrementing the variable~$\ell$ during the marching (line~\ref{alg2:inc}). In Algorithm~\ref{alg:extractGraph} each edge is traced twice which is redundant and could be improved. 

\begin{algorithm}[t]
\caption{ExtractGraph \label{alg:extractGraph}}
\KwIn{$\set V_s $, $\set V_t $  \tcp*{voxel skeleton and tagged voxels}}
\tcp{initialise graph $\set G$ with nodes~$\set P$, edges~$\set E$}
\tcp{ and function~$w$ mapping edges to weights}
$\set G = (\set P, \, \set E, \, w ) $  	\\
$\set P \leftarrow \emptyset; \,  \set E \leftarrow \emptyset$ \\
\For{$v\in\set V_s$}{
	\If{ {\normalfont jointOrTagged($v$)}  }{
		 $\set P \leftarrow \set P \cup \{ v\} $  \tcp*{edge starting at joint $v$}  		 
		 \For(\tcp*[f]{trace voxel chain}  \label{alg2:march}){$s \in \set N_{26} (v)$}{
		 	$ \ell = 2 ; \, s_{prev} = v $ \\			
			\While{  {\normalfont ( \textbf{not} jointOrTagged($s$) )}   }{
				$ s_{next} =  \set N_{26}(s) \setminus \{ s_{prev} \}$ \\ 
				$ s_{prev} =  s; \,  s =  s_{next}$ \\
				$\ell \leftarrow \ell + 1 $ \label{alg2:inc}
			}					
			$\set P  \leftarrow \set P \cup \{ s\} $  \tcp*{edge ending at joint $s$}   
			$\set E  \leftarrow \set E \cup \{ \{v, \, s \} \} $ \\ 
			$w (  \{v, \, s \} )  \leftarrow \ell $
		}	
	}			
}
\KwOut{$\set G = (\set P, \, \set E, w ) $  \tcp*{weighted undirected graph}}
\end{algorithm}

\clearpage
\bibliographystyle{elsarticle-num-names}
\bibliography{cadOpt}

\end{document}